\documentclass{IEEEtran}
\usepackage[T1]{fontenc}
\usepackage{textcomp}
\usepackage{graphicx}
\usepackage{todonotes}
\usepackage{amssymb,amsmath,amsthm}
\usepackage{tikz,pgfplots}
\usepackage{bigstrut}
\pgfplotsset{compat=newest}
\usepackage{algpseudocode}
\usepackage{algorithm}
\usepackage{fancyhdr}

\newcommand{\user}{\mathrm{user}}
\newcommand{\iter}{\mathrm{iter}}
\newcommand{\proc}{\mathrm{proc}}
\newcommand{\lat}{\mathrm{lat}}
\AtBeginDocument{
  \addtolength{\abovedisplayskip}{-5pt}
  \addtolength{\abovedisplayshortskip}{-5pt}
  \addtolength{\belowdisplayskip}{-5pt}
  \addtolength{\belowdisplayshortskip}{-5pt}
  \addtolength{\textfloatsep}{-5pt}
  \addtolength{\dblfloatsep}{-5pt}
  \addtolength{\abovecaptionskip}{-7pt}
  \addtolength{\belowcaptionskip}{-7pt}
}

\begin{document}

\title{Impact of Processing-Resource Sharing on the Placement of Chained Virtual Network Functions}

\author{
\IEEEauthorblockN{Marco Savi, Massimo Tornatore, Giacomo Verticale\\
\thanks{Marco Savi is with Fondazione Bruno Kessler, CREATE-NET Research Center, 38123, Trento, Italy (email: m.savi@fbk.eu).}
\thanks{Massimo Tornatore and Giacomo Verticale are with the Dipartimento di Elettronica, Informazione e   Bioingegneria (DEIB), Politecnico di Milano, 20133, Milan, Italy.}
\thanks{A preliminary version of this paper appeared in \cite{savi15}.}
}
}
\maketitle

\thispagestyle{fancy}
\renewcommand{\headrulewidth}{0pt}
\chead{\scriptsize This is the authors' version of an article that has been accepted for publication in IEEE Transactions on Cloud Computing. Changes were made to\\this version by the publisher prior to publication. The final version of record is available at {\color{blue}http://dx.doi.org/10.1109/TCC.2019.2914387}}

\cfoot{\scriptsize Copyright (c) 2019 IEEE. Personal use is permitted. For any other purposes, permission must be obtained from the IEEE by emailing pubs-permissions@ieee.org.}

\begin{abstract}
Network Function Virtualization (NFV) provides higher flexibility for network operators and reduces the complexity in network service deployment.
Using NFV, Virtual Network Functions (VNF) can be located in various network nodes and chained together in a Service Function Chain (SFC) to provide a specific service.
Consolidating multiple VNFs in a smaller number of locations would allow decreasing capital expenditures. However, excessive consolidation of VNFs might cause additional latency penalties due to processing-resource sharing, and this is undesirable, as SFCs are bounded by service-specific latency requirements.
In this paper, we identify two different types of penalties (referred as \lq \lq costs'') related to the processing-resource sharing among multiple VNFs: the {\em context switching costs} and the {\em upscaling costs}. Context switching costs arise when multiple CPU processes (e.g., supporting different VNFs) share the same CPU and thus repeated loading/saving of their context is required. Upscaling costs are incurred by VNFs requiring multi-core implementations, since they suffer a penalty due to the load-balancing needs among CPU cores.
These costs affect how the chained VNFs are placed in the network to meet the performance requirement of the SFCs. 
We evaluate their impact while considering SFCs with different bandwidth and latency requirements in a scenario of VNF consolidation.
\end{abstract}

\section{Introduction}
In the last years, \emph{Network Function Virtualization} (NFV)
has emerged as a promising technique to help network operators reduce capital and energy costs.
NFV is based on the concept of \emph{network function}, which is an abstract building block representing a piece of software designed to process the network traffic and accomplish a specific task.
Examples of network functions are firewalls, network address translators, traffic monitors, or even more complex entities such as 4G/5G service or packet gateways.
So far, network functions have been implemented using dedicated hardware referred to as \emph{middleboxes}. Such middleboxes are able to handle heavy traffic loads, but have an expensive and slow provisioning cycle. Additionally, they cannot be easily re-purposed and must be dimensioned at peak loads, leading to waste of resources when the traffic is low, e.g., in off-peak hours.
The NFV paradigm consists in moving from a hardware to a software implementation of network functions in a virtualized environment. This way, multiple and heterogeneous \emph{virtual network functions} (VNFs) can be hosted by the same generic commercial-off-the shelf (COTS) hardware. NFV adds flexibility to the network since it allows network operators to efficiently consolidate the VNFs and makes it possible on-the-fly provisioning.
Another value added by NFV is the simplicity in the deployment of heterogeneous network services. NFV exploits the concept of \emph{service function chaining} \cite{service_chaining}, according to which a service (e.g., web browsing, VoIP, etc.) can be provided by one or more service function chains (SFCs), i.e., a concatenation of appropriate VNFs that must be crossed by the traffic associated to that specific service.
The main weakness of NFV is the hard-to-predict performance due to resource sharing of hardware among different functions, especially concerning the \emph{processing} \cite{zeng18}.

In this paper, we evaluate the impact of processing-resource sharing on the placement of VNFs and on the embedding of SFCs in a VNF consolidation scenario, i.e., when we want to minimize the amount of COTS hardware deployed in the network.
We identify two sources of inefficiency and performance degradation. 
The first, which we will refer to as \emph{context switching costs}, stems from the need of sharing Central Processing Unit (CPU) resources among different VNFs and results in additional latency, since packets \emph{(i)} must wait for the correct VNF to be scheduled and \emph{(ii)} some CPU time is wasted due to the need of saving/loading the state of the VNFs at each scheduling period.
The second, which we will refer to as \emph{upscaling costs},
represents the additional latency and processing cost \emph{(i)} of balancing network traffic among multiple CPU cores in multi-core architectures and \emph{(ii)} of keeping the shared state synchronized among the NFV instances running over different cores.

Such performance degradation affects how the VNFs must be placed in the network to guarantee the requirements for different types of SFCs. Specifically, we propose a novel detailed node model that takes into account the aforementioned processing-resource sharing costs. Compared to existing models where resource-sharing penalties are not considered (e.g. \cite{chua16}), our model enables a more accurate distribution and scaling of VNFs, preventing excessive VNF consolidation when it might jeopardize SFC performance. To the best of our knowledge, this paper is the first study evaluating the impact of such processing-resource sharing costs on service function chaining in an NFV scenario. 

The remainder of the paper is organized as follows. Section \ref{sec:related_works} discusses related work concerning both processing-resource sharing and VNF placement. In Section \ref{sec:system_model} we introduce our system model by modeling the physical network, the VNFs, the SFCs and the processing-resource sharing costs. Section \ref{sec:vnf_consolidation} introduces the problem of VNF consolidation. We formulate an Integer Linear Programming (ILP) model and propose a heuristic algorithm to solve the defined problem in a scalable way. In Section \ref{sec:results} we show illustrative numerical results over a realistic network scenario. We first compare results obtained by solving the ILP model and by running our heuristic algorithm, then move our focus to the embedding of a more diverse set of SFCs and to the comparison with the state of the art. Finally, Section \ref{sec:conclusion} draws the conclusion of our work.

\section{Related work} \label{sec:related_works}
Our work is related both to processing-resource sharing in multi-core architectures and to VNF placement/SFC embedding. In the next two subsections we recall the related work with respect to such topics. 

\subsection{Processing-resource sharing}
Several studies in literature have investigated the challenges arising from processing-resource sharing. Refs. \cite{mccool08}\cite{patterson10}\cite{Fusco10} were among the first works investigating processing-resource sharing challenges due to the adoption of multi-core architectures. Refs. \cite{mccool08}\cite{patterson10} survey the architectural upgrades needed to efficiently scale processing performance by adopting multi-core technologies. Ref. \cite{Fusco10} argues that, even if the adoption of multi-core systems is the dominant trend, network devices hardly fully exploit multiple cores.
Among the challenges related to multi-core systems, \emph{load balancing} is one of the most complex. For example, Ref. \cite{load_balancing_multi_core} investigates how load balancing, by adding a new layer in the system architecture that can become a bottleneck, must be carefully designed and could lead to performance penalties, while Ref. \cite{Buh14} defines a novel adaptive traffic distribution among the CPU cores on a per-packet basis  trying to mitigate such issue. As we will better describe later, we call \emph{upscaling costs} the performance degradation due to load balancing.

Due to processing-resource sharing, another issue arises related to a well-known operation performed by processors, called \emph{context switching}. Context switching has been thoroughly investigated in literature and it is related to the need of saving/loading the context (i.e., the state) of a CPU process to enable the execution of multiple CPU processes on a single CPU. Ref. \cite{context_switch} defines a methodology to quantify the costs related to context switching in terms of latency, while Ref. \cite{Fromm_revisitingthe} investigates context switching costs due to cache interference among multiple CPU processes. Additionally, Ref. \cite{zhao18} focuses on analogous issues but related to accelerated services provided by Graphics Processing Units (GPUs). Refs. \cite{asai14}\cite{cerrato_vnf_1}\cite{cerrato_vnf_2} have instead investigated the impact of context switching on NFV. Specifically, Ref. \cite{asai14} defines some strategies to reduce the context switching costs, while Refs. \cite{cerrato_vnf_1}\cite{cerrato_vnf_2} design and implement algorithms for efficient sharing of processing resources among VNFs, which are considered as specific types of CPU processes.
Finally, Ref. \cite{Li17} presents a latency-aware NFV scheme where software middleboxes can be dynamically scheduled  according to the changing traffic and resources, which affect latency in the network due to processing-resource sharing. 
 In our paper we also deal with upscaling and context switching costs, but our goal is evaluating their impact on VNF placement and SFC embedding.

\subsection{VNF placement and SFC embedding}
We formalize an optimization problem for VNF placement and SFC embedding that can be seen as an extension of some well-known Virtual Network Embedding (VNE) problems, as the ones shown in Refs. \cite{vne_botero}\cite{vne_fuerst}\cite{vne_chowdhury}. In our problem, the SFCs are the virtual networks and the chained VNFs are virtual nodes, which are connected by virtual links and must be crossed in sequential order. Such SFCs must be embedded in a physical network, where each virtual link can be mapped to a physical path \cite{vne_botero}, multiple virtual nodes can be mapped to the same physical node \cite{vne_fuerst}, and virtual nodes must be consolidated \cite{vne_chowdhury}. In our SFC embedding problem it must be also guaranteed that a virtual node  can be shared among multiple virtual networks. From another point of view, the SFCs can be seen as \emph{walks} on the physical graph.
Ref. \cite{mehraghdam14} is the first work investigating the optimal embedding of SFCs in the network following a VNE approach. The authors formulate a Mixed Integer Quadratically Constrained Problem to evaluate the optimal placement of VNFs. In our paper, we formulate a similar problem, but we extend the analysis to cover also processing-resource sharing aspects. 

Some other studies have dealt with the placement of VNFs in the network. Refs. \cite{moens14}\cite{addis15}\cite{gupta15}\cite{mohammadkhan15}\cite{bari15}\cite{bouet15}\cite{riggio15_1}\cite{Carpio16}\cite{ren18}\cite{guo18} all formulate an ILP model to solve the problem of optimal VNF placement and/or SFC embedding, considering different objective functions. Ref. \cite{moens14} minimizes the used servers, Refs. \cite{addis15}\cite{mohammadkhan15}\cite{Carpio16} the maximum utilization of links, Ref. \cite{bari15} the OPEX and resource utilization, Ref. \cite{bouet15} the network load, Ref. \cite{riggio15_1} the VNF mapping cost, Ref. \cite{gupta15} the overall consumed bandwidth and Ref. \cite{ren18} the traffic delivery cost. Instead, Ref. \cite{guo18} maximizes the throughput. Except for Refs. \cite{moens14}\cite{addis15}\cite{gupta15}, all the other works also present a heuristic algorithm to improve the scalability of the model. 
Concerning about heuristic algorithms for VNF placement, Ref. \cite{cao16} proposes four genetic algorithms for network load balancing, while Ref. \cite{hirwe16} aims at minimizing the SFC embedding cost. In our work the objective is different, as we want to maximally consolidate the VNFs as Ref. \cite{moens14}, by taking into account processing-resource sharing, which adds more practical flavor to the solution of the problem.

While solving an ILP allows to obtain a solution for a static placement of VNFs, some other papers deal with the definition and implementation of online algorithms for an on-the-fly and dynamic deployment. Specifically, Refs. \cite{callegati15}\cite{oechsner15} implement dynamic VNF chaining by means of OpenFlow \cite{callegati15} and OpenStack \cite{oechsner15}.
Refs. \cite{riera14}\cite{Ghribi16}\cite{mijumbi15} define some algorithms for online VNF scheduling and placement: while Refs. \cite{riera14}\cite{Ghribi16} take into account service function chaining aspects, Ref. \cite{mijumbi15} does not. Refs. \cite{sahhaf15}\cite{lukovszki15}\cite{ghaznavi15}\cite{fei18} focus on the definition of algorithms for online embedding of SFCs while minimizing the resources consumed by the infrastructure to embed each SFC request. Unlike the other works, Ref. \cite{fei18} also leverages an online learning method for service demand prediction and proactive VNF provisioning. Finally, Ref. \cite{guo18} designs an online algorithm with proven competitive ratio with respect to the objective of maximizing throughput. Even though in our paper we do not design and implement any online algorithm for SFC embedding, we define a heuristic algorithm for VNF consolidation that embeds the SFC one-by-one and that can be used as a basic engine for any online algorithm.

\section{System model} \label{sec:system_model}

Our model of NFV-enabled network comprises representations of: an Internet Service Provider (ISP) network with physical links and nodes, a set of edge-to-edge\footnote{In this paper we deal with the embedding of aggregated SFCs between the edges of the core network (i.e., \emph{ISP network}), serving multiple users belonging to the same access/aggregation networks. For this reason, we will use the terms \emph{end-to-end} and \emph{edge-to-edge} interchangeably.} services implemented as a chain of  VNFs, a set of atomic VNFs that can be deployed in an NFV-capable physical node.
Our model of NFV nodes is in line with current trends towards the re-architecture of core nodes as \emph{micro-datacenter} able to host several VNFs \cite{peterson16}: such NFV nodes have a limited amount of processing power, which is split over multiple computational cores. To achieve parallelism, any VNFs must adopt load balancing mechanisms, resulting in the employment of CPU time and in additional latency to traverse the node. Furthermore, multiple VNFs can be involved in computational power contention, resulting in the employment of CPU time for VNF coordination and in additional latency for the packets that must wait the scheduling of their reference VNF.

\subsection{Physical network and SFC/VNF modeling}
\subsubsection{Physical network}

\begin{figure}[t]
\centering
\includegraphics[width=1\columnwidth]{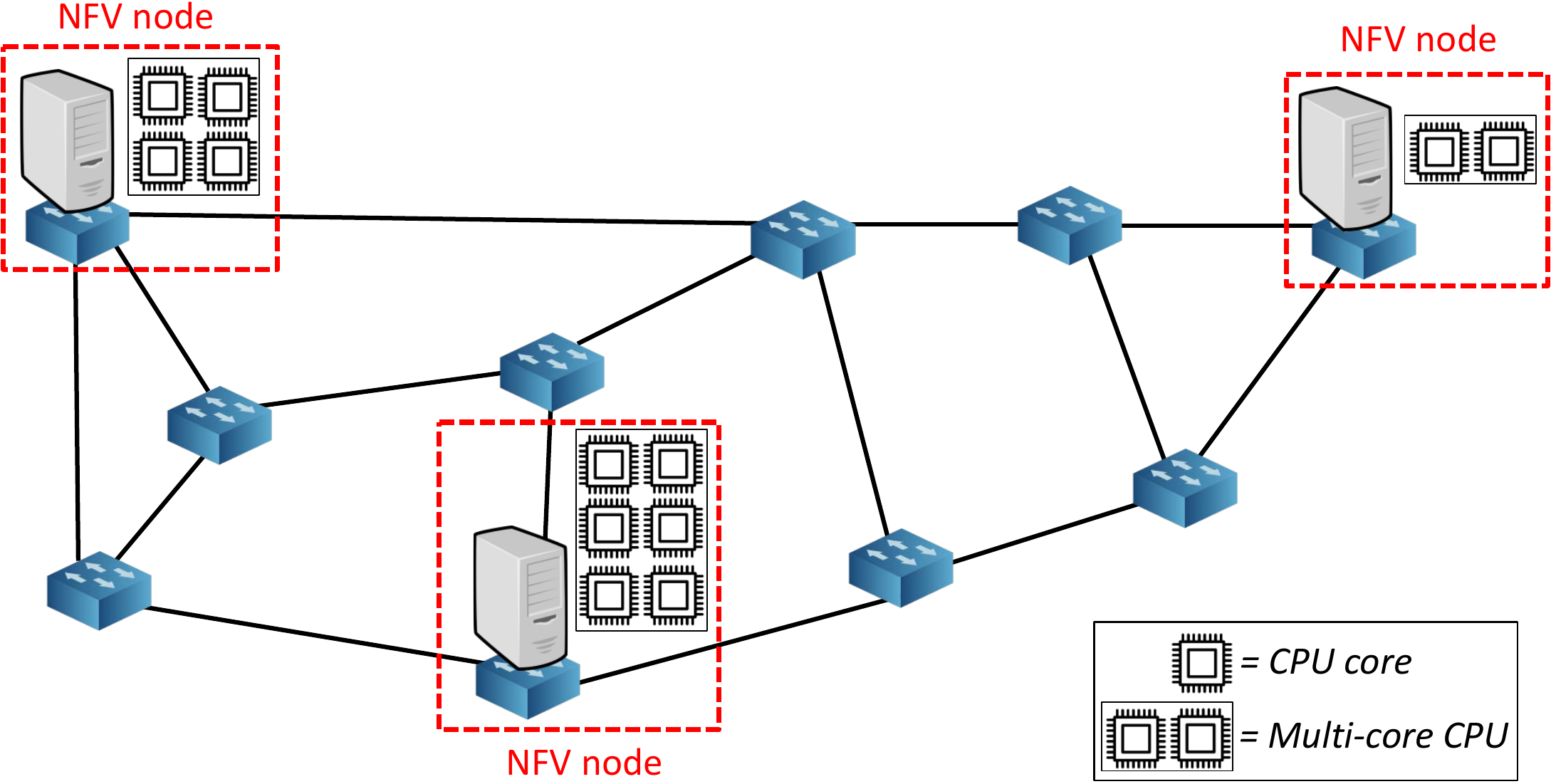}\caption{Physical topology where some nodes are equipped with generic multi-core COTS hardware (i.e., \emph{NFV nodes}) \label{fig:topology}}
\end{figure}

We model the physical network as a connected directed graph $\mathcal{G}=(V,E)$.
All the network nodes $v \in V$  have basic packet forwarding capabilities.
The links $(v,v')\in E$ have capacity $\beta_{v,v'}$ and latency $\lambda_{v,v'}$.
A subset of the network nodes $v$ can be also equipped with hardware capable of executing VNFs.
The model is agnostic with respect to the physical location of such nodes, which can be cabinets, central offices, core exchanges, etc.
We generally refer to this nodes as \emph{NFV nodes} (see Fig. \ref{fig:topology}). Every NFV node is connected to an ideal zero-latency ($\lambda_{v,v}=0$) and infinite-bandwidth ($\beta_{v,v}=\infty$) self-loop $(v,v)\in E$. Its use will be discussed later in the paper.
An NFV node is also equipped with a \emph{multi-core CPU}. We measure its processing capacity $\gamma_v$, expressed in terms of number of CPU cores that the multi-core CPU supports.
If a physical node has only forwarding capabilities (i.e., it is a forwarding-only node, such as a legacy router or a switch), it follows that $\gamma_v=0$.
In the rest of the paper we will assume a one-to-one correspondence between an NFV node and a multi-core CPU, and we will interchangeably use the terms \emph{multi-core CPU} and \emph{NFV node}.

\subsubsection{Service function chains}
When a network operator provisions a \emph{service} between two end-points, it deploys one or more SFCs $c$ in the network. For the sake of simplicity, we consider a one-to-one correspondence between a service and an SFC: thus, the carrier's goal is to embed a set of SFCs $C$.

A single SFC $c$ can be modeled by a walk
$\mathcal{C}^c=(X^c \cup U^c, G^c)$,
where $X^c$ is the set of start/end points $u$,
$U^c$ is the set of VNF requests $u$ and
$G^c$ is the set of virtual links $(u,u')$ chaining consecutive VNF requests/start/end points $u$ and $u'$.

\begin{figure}[t]
\centering
\includegraphics[width=1\columnwidth]{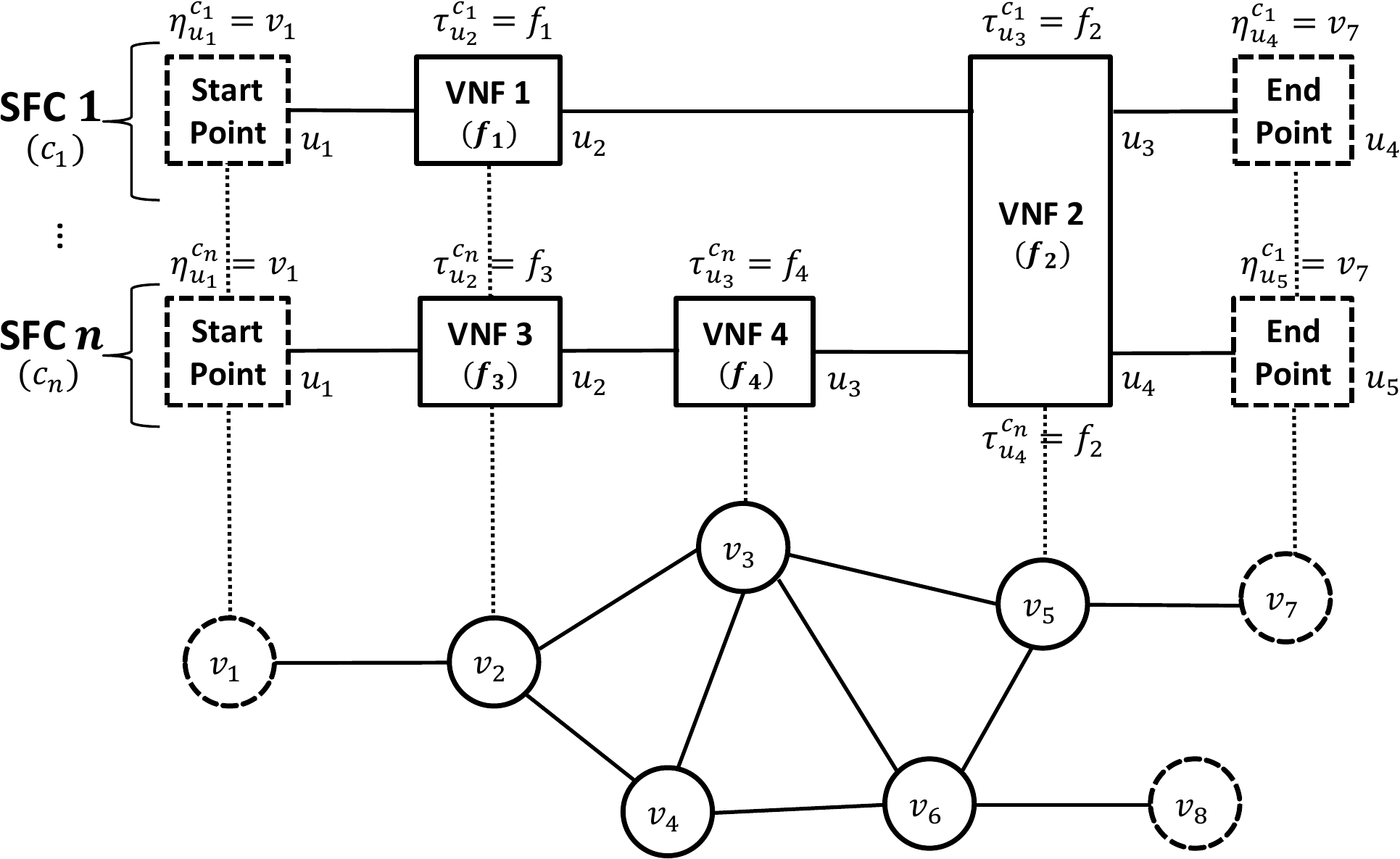}\caption{Example of SFCs that must be embedded in the physical network, where different VNFs can share the processing resources of the same NFV node and multiple SFCs can share the same VNF \label{fig:vne}}
\end{figure}

From a topological perspective, both VNF requests and start/end points are virtual nodes
$u \in X^c \cup U^c$.
Note that in our model, as done in \cite{mehraghdam14}, we decouple the concepts of
\emph{VNF} $f\in F$ and of \emph{VNF request} $u\in U^c$.
For each SFC $c$, we have a chain of VNF requests $u$; each VNF request $u$
is mapped to a specific VNF $f$ through the mapping parameter
$\tau^c_u=f \in F$.
The input parameter $\tau^c_u$ allows us to relate each VNF request $u$ to the specific VNF $f$ it requests (see Fig. \ref{fig:vne}). Moreover, what we place in the NFV nodes are \emph{VNF instances} of different VNFs $f$. In this way, multiple VNF requests $u$ for different SFCs $c$ can be mapped to the same VNF instance of a VNF $f$.
Similar considerations can be done for the start/end points $u\in X^c$.
Such points are fixed in the network, and we introduce the input mapping parameter $\eta_u^c = v\in V$ to make explicit the mapping between the start/end point $u$ for the SFC $c$ to a specific physical node $v$ (see Fig. \ref{fig:vne}).

In our model, every SFC $c$ can serve an aggregated \emph{number of users} $N^{\user}$. Such aggregated SFCs are deployed when multiple users require the same service between the same pair of start/end point. Every SFC is then associated to a set of performance constraints:
\begin{itemize}
\item The \emph{aggregated requested bandwidth} $\delta_{u,u'}^c$, i.e., the bandwidth that must be guaranteed between two VNF requests/start/end points $u$ and $u'$ to support the service offered by the SFC $c$ for all the users. It follows that $\delta^c_{u,u'} = N^{\user} \delta_{u,u'}^{c,\user}$, where $\delta_{u,u'}^{c,\user}$ is the \emph{requested bandwidth per user} for each virtual link $(u,u')$. Every virtual link $(u,u')$ can be associated to a different bandwidth requirement since the chained VNFs can lead to a change in the traffic throughput.
\item The \emph{maximum tolerated latency} $\varphi^c$, i.e., the maximum end-to-end delay that can be introduced by the network without affecting the service between the start/end points of the SFC $c$.
\end{itemize}

\subsubsection{Virtual network functions}
A VNF can be seen as a black-box performing some operations on the network traffic.
Every VNF $f$ hosted by an NFV node $v$ (i.e., every VNF instance) must be assigned a dedicated amount of processing $c_{f,v}$ per time unit in order to perform all the necessary operations on the input traffic. The processing $c_{f,v}$ is expressed as a number or fraction of the CPU cores $\gamma_v$ of the NFV node $v$ assigned to the instance of VNF $f$. For example, $c_{f,v}=2.5$  means that the instance of VNF $f$ fully consumes, on the NFV node $v$, the processing resources of two CPU cores plus half the resources of a third CPU core. We say that a VNF instance has a larger \emph{size} when it is assigned a larger processing capability $c_{f,v}$.
We then define $\pi_f$ as the \emph{processing per VNF request}, also expressed as fraction of CPU cores.
$\pi_f $ indicates the processing resources that must be dedicated to each VNF request $u$ mapped to an instance of VNF $f$. The ratio $c_{f,v}/\pi_f$ is thus the theoretical maximum number of VNF requests that can share the instance of VNF $f$ hosted by the NFV node $v$. $\pi_f$ depends on the number of users $N^{\user}$ per each SFC $c$: in general, we have that $\pi_f = N^{\user} \pi_f^{\user}$, where $\pi_f^{\user}$ is the \emph{processing per user} for the VNF $f$.
Note that different VNFs are characterized by a $\pi_f^{\user}$ that can largely vary from one to each other, depending on the complexity of the performed operations.

\subsection{Modeling of processing-resource sharing costs} \label{sec:model_processing_resource_sharing}
\begin{figure}[t]
	\centering
	\includegraphics[width=1\columnwidth]{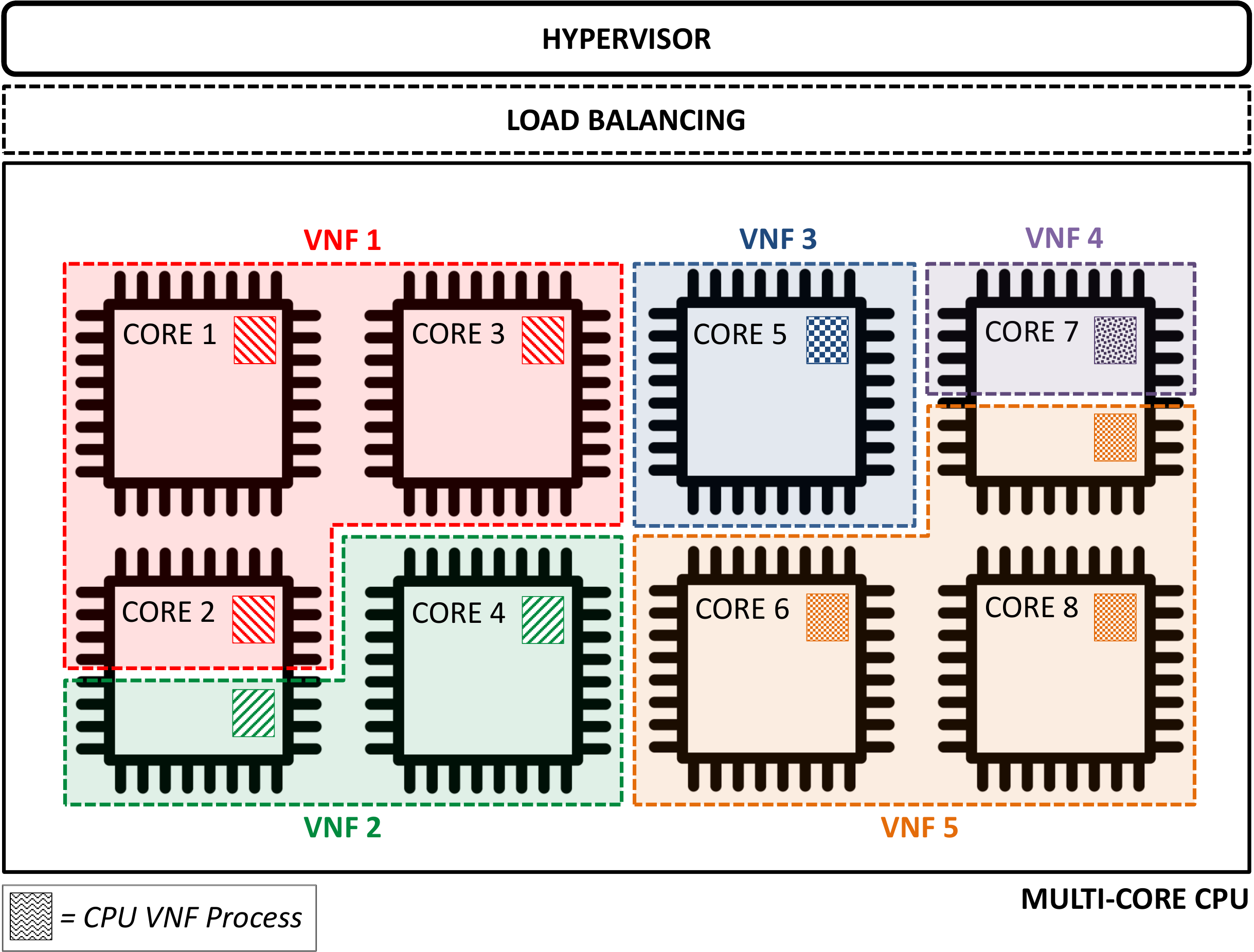}\caption{Pictorial view of CPU sharing among multiple VNFs, responsible for the \emph{context switching costs} \label{fig:context_switching_costs}}
\end{figure}

As seen in Section \ref{sec:related_works}, when multiple VNFs share the computational resources available at the same NFV node, some performance degradation due to processing-resource sharing is expected. As already introduced, we have identified two types of costs:  \emph{context switching costs} and \emph{upscaling costs}.

\subsubsection{Context switching costs}
\begin{figure}[t]
	\centering
	\includegraphics[width=1\columnwidth]{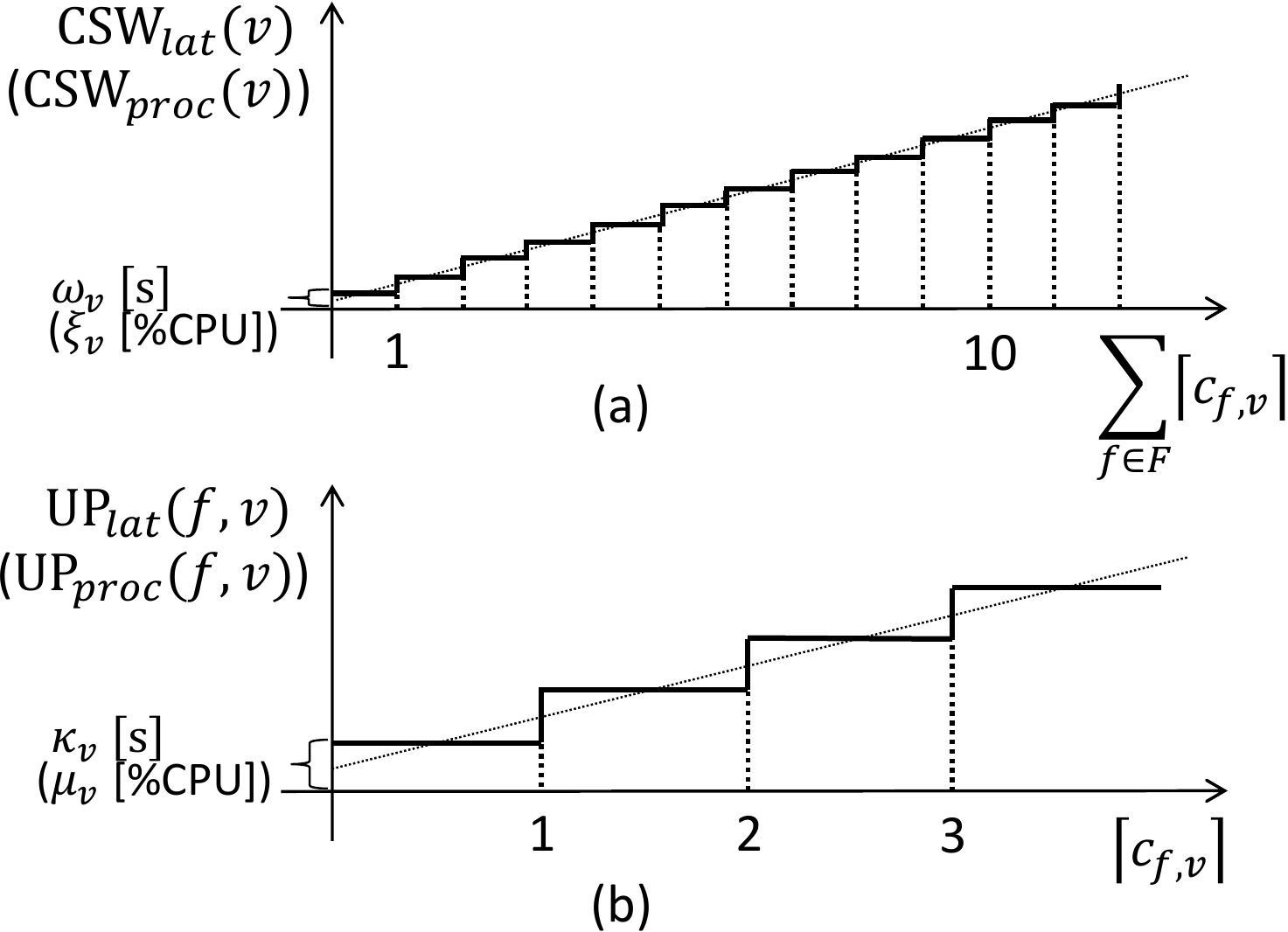}\caption{Modelling of context switching (a) and upscaling (b) costs \label{fig:up_csw}}
\end{figure}

In an NFV node $v$ the CPU cores are shared among instances of different VNFs $f$. The processing resources of a single core can be used by a VNF through a dedicated \emph{process}.
Fig. \ref{fig:context_switching_costs} shows an example where the processing resources of an 8-core CPU are shared among six different VNFs. Every time a VNF requires processing resources by multiple cores, a different process used by that VNF must be executed on each core.
Having multiple processes sharing multiple cores leads to performance penalties which we will refer to as \emph{context switching} costs.
In fact, the CPU needs some time and some dedicated processing capacity to perform the operation of context switching. The degradation effects due to context switching are then an increase in the \emph{latency} introduced by the NFV node (i.e., latency costs) and a reduction of the \emph{actual processing capacity} that can be used to host the VNFs (i.e., processing costs). According to \cite{cerrato_vnf_2}, we model the context switching costs as linearly increasing with respect to the overall number of processes sharing the NFV node $v$, as shown in Fig. \ref{fig:up_csw}(a).

We can express such costs in the following way:
\begin{equation}
\text{CSW}_{\lat}(v) = \sum_{f\in F} \lceil c_{f,v} \rceil \omega_v
\end{equation}
\begin{equation}
\text{CSW}_{\proc}(v) = \sum_{f\in F} \lceil c_{f,v} \rceil \xi_v
\end{equation}
where $\sum_{f\in F} \lceil c_{f,v} \rceil$ indicates the overall number of processes involved in the context switching operation for the NFV node $v$, $\omega_v$ is the context switching latency parameter (measured in time units) and $\xi_v$ is the context switching processing parameter (measured in number or fractions of CPU cores). Such parameters can vary for different NFV nodes depending on the adopted multi-core CPU technology. If follows that $\text{CSW}_{\lat}(v)$ is a time quantity (e.g. expressed in ms) and $\text{CSW}_{\proc}(v)$ is a fraction of the CPU cores.

\subsubsection{Upscaling costs}
\begin{figure}[t]
	\centering
	\includegraphics[width=1\columnwidth]{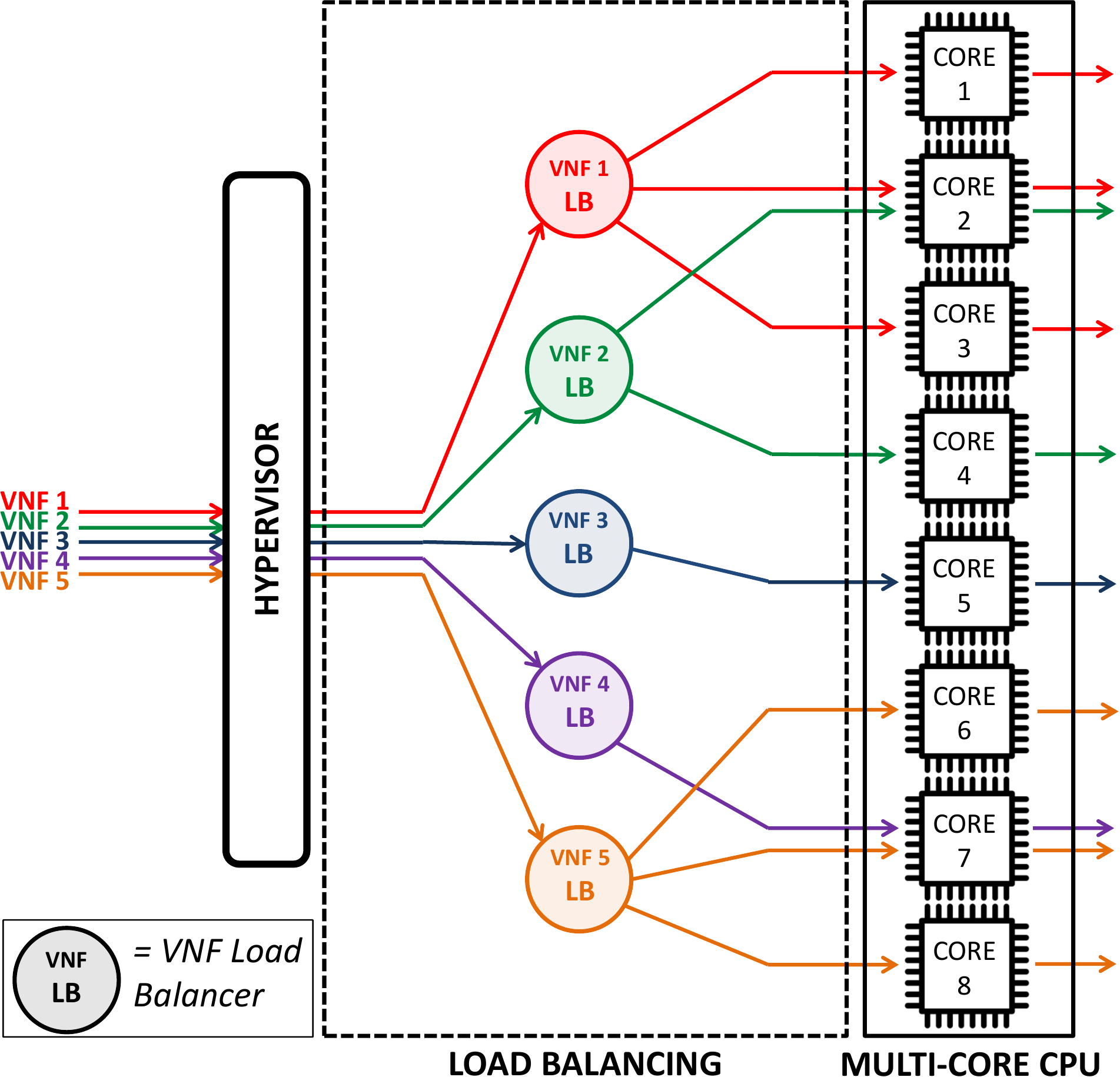}\caption{Pictorial view of the load balancing layer, responsible for the \emph{upscaling costs} \label{fig:upscaling_costs}}
\end{figure}
As stated above, an instance of a VNF $f$ placed on an NFV node $v$ can require more than the processing resources provided by a single CPU core. This happens when the VNF must process a high quantity of traffic (e.g., when a high number of users shares such VNF). In this case, the network traffic handled by that VNF must be balanced among multiple CPU cores. Figure \ref{fig:upscaling_costs} shows how network traffic must be balanced according to the example shown in Fig. \ref{fig:context_switching_costs}. The new layer of \emph{load balancing} shown in Fig. \ref{fig:upscaling_costs} is responsible for the \emph{upscaling costs}. In fact, every VNF that is hosted by the NFV node needs a dedicated \emph{load balancer} that takes the decision on how the traffic is sorted among the CPU cores.
The load balancer can be itself seen as an auxiliary VNF performing the specific task of balancing traffic among the CPU cores: it thus needs some time to take the balancing decision and some dedicated processing capacity to perform such operation. As for context switching, the overall degradation effects due to load balancing are an increase in the \emph{latency} introduced by the NFV node for the considered VNF (i.e, latency costs) and a reduction of the \emph{actual processing capacity} that can be used to host the VNFs (i.e., processing costs).
We model the upscaling costs, both concerning latency and processing, as linearly increasing with respect to the number of CPU cores among which the traffic is balanced, as shown in Fig. \ref{fig:up_csw}(b). We can express such costs as:
\begin{equation}
\text{UP}_{\lat}(f,v) = \lceil c_{f,v} \rceil \kappa_v
\end{equation}
\begin{equation}
\text{UP}_{\proc}(f,v) = \lceil c_{f,v} \rceil \mu_v
\end{equation}
where $\lceil c_{f,v} \rceil$ indicates the number of cores involved in the load balancing for the instance of VNF $f$ placed in NFV node $v$. $\kappa_v$ is the upscaling latency parameter and $\mu_v$ is the upscaling processing parameter. They can differ for heterogeneous NFV nodes, depending on how the load balancing layer is implemented. If follows that $\text{UP}_{\lat}(f,v)$ is a time quantity and $\text{UP}_{\proc}(f,v)$ is a fraction of CPU cores.

Note that, in this paper, we focus on how physical processing resources are shared among different VNFs in an ISP network with micro-datacenters, where no horizontal VNF \emph{scale out} is expected (contrarily to what happens in large datacenters \cite{chi15}). How virtual processing resources (i.e., vCPUs) are mapped to physical resources (i.e., CPU cores) is a task performed by a \emph{Hypervisor}, depicted in Fig. \ref{fig:context_switching_costs} and Fig. \ref{fig:upscaling_costs}. Different hypervisors lead to a different mapping among physical and virtual resources. The evaluation of the performance of different hypervisors is outside the scope of this paper, in which we assume that an optimal mapping among physical and virtual resources always occurs.

\begin{table}[t]
	\centering
	\caption{Summary of graphs and sets considered in the model} \label{tab:sets}
	\begin{tabular}{|p{2.55cm}|p{5.2cm}|}
		\hline
		\centering Graph/Set & Description\\
		\hline
		\centering $\mathcal{G}=(V,E)$ & Physical network graph, where $V$ is the set of physical nodes $v$ and $E$ is the set of physical links $(v,v')$ connecting the nodes $v$ and $v'$\\
		\hline
		\centering $C$ & Set of the service function chains $c$ that must be embedded in the physical network $\mathcal{G}$\\
		\hline
		\centering $\mathcal{C}^c=(X^c\cup U^c,G^c)$ & Virtual graph for the SFC $c$, where $X^c$ is the set of fixed start/end point $u$, $U^c$ is the set of VNF requests $u$, $G^c$ is the set of virtual links $(u,u')$ connecting the VNF request (or start point) $u$ and the VNF request (or end point) $u'$\\
		\hline
		\centering $F$ & Set of VNFs $f$ that can be requested and placed in the network\\
		\hline
	\end{tabular}
\end{table}
\begin{table} [t]
	\centering
	\caption{Summary of parameters considered in the model} \label{tab:parameters}
	\begin{tabular}{|c|p{1.7cm}|p{4.5cm}|}
		\hline \centering
		Parameter & Domain & Description\\
		\hline
		\centering $\tau^c_u$ & $c \in C$\newline $u \in U^c$ & VNF requested by the VNF request $u$ in the SFC $c$ ($\tau^c_u \in F$)\\
		\hline
		\centering $\eta^c_u$ & $c\in C$\newline $u \in X^c$ & Physical node where the start/end point $u$ for the SFC $c$ is mapped ($\eta^c_u \in V$)\\
		\hline
		\centering $\gamma_v$ & $v \in V$ & Number of the CPU cores (i.e., processing capacity) hosted by the node $v$ \\
		\hline
		\centering $\beta_{v,v'}$ & $(v,v')\in E$ & Bandwidth capacity of the physical link $(v,v')$\\
		\hline
		\centering $\lambda_{v,v'}$ & $(v,v')\in E$ & Latency of the physical link $(v,v')$\\
		\hline
		\centering $\omega_v$ & $v\in V$ & Context-switching latency of the node~$v$\\
		\hline
		\centering $\xi_v$ & $v \in V$ & Context-switching processing of the node $v$\\
		\hline
		\centering $\kappa_v$ & $v\in V$ & Upscaling latency of the node $v$\\
		\hline
		\centering $\mu_v$ & $v\in V$ & Upscaling processing of the node $v$\\
		\hline
		\centering $N^{\user}$ & & Number of aggregated users per SFC\\
		\hline
		\centering $\delta^{c,\user}_{u,u'}$ &  $c\in C\newline (u,u')\in G^c$ & Requested bandwidth per user on the virtual link $(u,u')$ of the SFC $c$\\
		\hline
		\centering $\delta^{c}_{u,u'}$ &  $c\in C\newline (u,u')\in G^c$ & $= N^{\user} \delta_{u,u'}^{c,\user}$: Requested bandwidth on the virtual link $(u,u')$ per SFC~$c$\\
		\hline
		\centering  $\varphi^c$ & $c\in C$ & Maximum tolerated latency by the SFC~$c$\\
		\hline
		\centering $\pi_f^{\user}$ & $f \in F$ & Fraction of the CPU core (i.e., processing requirement)  per user for the VNF~$f$\\
		\hline
		\centering $\pi_f$ & $f \in F$ & $= N^{\user} \pi_f^{\user}$: Fraction of the CPU processing required by each VNF request $u$ for the VNF $f$\\
		\hline
	\end{tabular}
\end{table}

\section{Solving the Problem of VNF Consolidation} \label{sec:vnf_consolidation}
A summary of the sets and parameters defined in Section \ref{sec:system_model} and used in this section is reported in Tables \ref{tab:sets} and \ref{tab:parameters}. In the following we focus on the problem of \emph{VNF consolidation}: given a physical network topology and some SFCs, we want to decide position and size, in terms of dedicated CPU cores, of the chained VNFs while minimizing the number of \emph{active} NFV nodes (i.e., the nodes hosting at least one VNF) in the network. This optimization problem can be useful for network operators to plan the best placement of the COTS hardware, since the number of active NFV nodes in the network is a measure of the \emph{cost for NFV implementation}: it follows that the cost is minimized when the VNFs are maximally consolidated.
We adopt two different approaches for solving the problem of VNF consolidation. First, we define an \emph{ILP model} that, once solved, allows to obtain an optimal solution for it. However, as specified in Section \ref{sec:related_works}, we deal with a virtual network embedding problem and it is well-known that virtual network embedding problems are NP-hard \cite{vne_chowdhury}. For this reason, we then design a heuristic algorithm called \emph{Heuristic Cost-aware Algorithm (HCA)} that allows to obtain a suboptimal solution in a much shorter time and could also be used for a real-time placement of SFCs.
\begin{table}
	\centering
	\caption{Decision variables for the ILP model} \label{tab:variables}
	\begin{tabular}{|c|p{1.6cm}|p{3.1cm}|}
		\hline \centering
		Variable & Domain & Description\\
		\hline
		\centering $m^c_{u,v} \in \{0,1\}$ & $c\in C\newline u\in U^c\newline v\in V$ & Binary variable such that $m^c_{u,v}=1$ iff the VNF request $u$ for the SFC $c$ is mapped to the node $v$, otherwise $m^c_{u,v}=~0$\\
		\hline
		\centering $c_{f,v} \in [0,\gamma_v]$ & $f\in F \newline v\in V$ & Real variable indicating the fraction of the CPU cores in the node $v$ used by the VNF $f$\\ 
		\hline
		\centering  $i_{f,v} \in \{0,1\}$ & $f\in F \newline v\in V$ & Binary variable such that $i_{f,v}=1$ iff the VNF $f$ is hosted by the node $v$, otherwise $i_{f,v}=0$\\
		\hline
		\centering $e^c_{v,v',x,y,u,u'} \in \{0,1\}$ & $c\in C \newline (v,v') \in E \newline x \in V \newline y \in V \newline (u,u')\in G^c$ & Binary variable such that $e^c_{v,v',x,y,u,u'}=1$ iff the physical link $(v,v')$ belongs to the path between the nodes $x$ and $y$, where the VNF requests $u$ and $u'$ for the SFC $c$ are mapped to, otherwise $e^c_{v,v',x,y,u,u'}~=0$\\
		\hline
		\centering $a_v \in \{0,1\}$ & $v\in V$ & Binary variable such that $a_v=1$ iff the node $v$ is active, otherwise $a_v =0$\\
		\hline
	\end{tabular}
\end{table}
\subsection{ILP model description} \label{sec:ilp_model}
In Table \ref{tab:variables} we have reported the decision variables for our ILP model. The model extends the formulation proposed in \cite{mehraghdam14} to consider, in the SFC embedding evaluation, the NFV node processing capacity and the VNF processing requirements (eq. \ref{eq:request_placement_2}-\ref{eq:c_i_relationship_2}). It also additionally includes upscaling and context switching costs in terms of added latency (eq. \ref{eq:latency_costs}-\ref{eq:latency}) and reduced node processing capacity (eq. \ref{eq:processing_costs}-\ref{eq:maximum_processing}), as modeled in Section \ref{sec:model_processing_resource_sharing}. In the following, we describe the objective function and the constraints. 

\subsubsection{Objective function}
\begin{equation}
\min \sum_{v\in V} a_v
\end{equation}
The objective function simply minimizes the number of active NFV nodes, as already proposed in \cite{mehraghdam14}.

\vspace{3pt}
The constraints are then grouped in three categories: \emph{request placement}, \emph{routing} and \emph{performance} constraints. The request placement constraints (eq. \ref{eq:fixed_point_1}-\ref{eq:maximum_processing}) ensure a correct mapping of VNFs to the NFV nodes as well as a correct mapping between VNF requests and VNFs. The routing constraints (eq. \ref{eq:e_m_relationship}-\ref{eq:no_loop_2}) guarantee a correct mapping of virtual links to physical paths. Finally, the performance constraints (eq. \ref{eq:bandwidth}-\ref{eq:user_indicator_2}) are related to performance requirements that must be guaranteed for both physical network and SFCs.

\subsubsection{Request placement constraints}
\begin{align} \label{eq:fixed_point_1}
 & \hspace{-30pt} m_{u,\eta^c_u}^c = 1 \quad c \in C, u\in X^c\\
& \hspace{-30pt} m^c_{u,v} = 0 \quad c \in C, u\in X^c, v \in V : v \neq \eta^c_u  \label{eq:fixed_point_2}
\end{align}
\begin{align} \label{eq:request_placement_1}
\hspace{-80pt}\sum_{v \in V} m_{u,v}^c=1 \quad c \in C, u\in U^c
\end{align}
\begin{align} \label{eq:request_placement_2}
\hspace{-78pt}\sum_{\substack{c\in C\\u\in U^c : \tau^c_u=f}} \hspace{-10pt} m_{u,v}^c \le  \frac{c_{f,v}}{\pi_f} \quad f \in F, v\in V
\end{align}
\begin{align} \label{eq:c_i_relationship_1}
& \hspace{-75pt}c_{f,v} \le \mathcal{M}_{\gamma} \ i_{f,v} \quad f \in F, v\in V\\
& \hspace{-75pt}i_{f,v} - c_{f,v} < 1 \quad f \in F, v\in V \label{eq:c_i_relationship_2}
\end{align}
\begin{align} \label{eq:one_vnf_mapping}
\hspace{-65pt} i_{f,v} \le \hspace{-10pt} \sum_{\substack{c\in C\\u\in U^c : \tau^c_u=f}} \hspace{-10pt} m_{u,v}^c \quad f \in F, v \in V
\end{align}
\begin{align} \label{eq:processing_costs}
\hspace{-30pt} \psi_v & =\text{CSW}_{\proc}(v) +  \sum_{f\in F} \text{UP}_{\proc}(f,v) \nonumber\\
 &  =\sum_{f\in F} \lceil c_{f,v}\rceil \xi_v + \sum_{f\in F} \lceil c_{f,v}\rceil  \mu_v \quad v \in V
\end{align}
\begin{align} \label{eq:maximum_processing}
\hspace{-85pt} \sum_{f\in F} c_{f,v} \le \gamma_v - \psi_v \quad v \in V
\end{align}
Eq. \ref{eq:fixed_point_1} guarantees that the fixed start/end point $u$ of a SFC $c$ is mapped to the node $v$ specified by the parameter  $\eta^c_u$ and eq. \ref{eq:fixed_point_2} that it is not mapped to any other node. Note that eq. \ref{eq:fixed_point_1} and eq. \ref{eq:fixed_point_2} fix the value for some variables, because start/end points of SFCs are a-priori known.

Every VNF request $u$ for each SFC $c$ must be mapped to exactly one node $v$ (eq. \ref{eq:request_placement_1}), in such a way that the overall number of VNF requests $u$ mapped to the VNF instance $f$ hosted by the node $v$ cannot overcome $c_{f,v}/\pi_f$ (eq. \ref{eq:request_placement_2}). Remind that $\pi_f = N^{\user} \pi_f^{\user}$, i.e., the higher the number of users per SFC, the higher the processing requirement per VNF.
Eq. \ref{eq:c_i_relationship_1} and eq. \ref{eq:c_i_relationship_2} ensure that $i_{f,v}=0$ if $c_{f,v}=0$ and that $i_{f,v}=1$ if $c_{f,v}>0$. $i_{f,v}$ is a flag variable specifying whether VNF $f$ is mapped to node $v$. $\mathcal{M}_{\gamma}$ is a big-M parameter, greater than the maximum value taken by $c_{f,v}$, i.e., $\mathcal{M}_{\gamma}>\max_{v \in V}\{\gamma_v\}$. Note that eq. \ref{eq:c_i_relationship_2} includes a strict inequality and must be linearized\footnote{All the linearization techniques used in this work follow standard approaches for linearization of constraints such as the ones shown in \cite{sherali2010reformulation}.}. Eq. \ref{eq:one_vnf_mapping} guarantees that an instance of VNF $f$ is placed on a node $v$ only if there is at least one request $u$ mapped to it.
Then, in eq. \ref{eq:processing_costs} we compute $\psi_v$, i.e., the overall context switching and upscaling processing costs per node $v$, as defined in Section \ref{sec:system_model}. Thus, the overall CPU processing assigned to instances of different VNFs $f$ cannot overcome the actual processing capacity $\gamma_v - \psi_v$ of node $v$ (eq. \ref{eq:maximum_processing}).
Note that the ceiling function $\lceil c_{f,v}\rceil$ that appears in eq. \ref{eq:processing_costs} must be linearized.

\subsubsection{Routing constraints}
\begin{align} \label{eq:e_m_relationship}
& e^c_{v,v',x,y,u,u'} \le m^c_{u,x}  \ m^c_{u',y} \nonumber\\
& c\in C, (v,v')\in E, x \in V, y \in V, (u,u')\in G^c
\end{align}
\begin{align} \label{eq:source}
&\hspace{-5pt}\sum_{\substack{(x,v)\in E, \ x,y\in V}} \hspace{-20pt}e^c_{x,v,x,y,u,u'}  \ m^c_{u,x} \ m^c_{u',y} = 1 \quad  c\in C, (u,u') \in G^c\\
&\hspace{-5pt}\sum_{\substack{(v,y)\in E, \ x,y\in V}} \hspace{-20pt}e^c_{v,y,x,y,u,u'} \ m^c_{u,x} \ m^c_{u',y} = 1 \quad  c \in C, (u,u') \in G^c \label{eq:destination}
\end{align}
\begin{align} \label{eq:source_2}
&\hspace{-55pt}\sum_{\substack{(v,x)\in E, \ v\in V}}\hspace{-15pt}e^c_{v,x,x,y,u,u'} = 0 \nonumber\\& \hspace{-33pt}c\in C, x\in V, y \in V, x\neq y,(u,u')\in G^c
\end{align}
\begin{align}
&\hspace{-55pt}\sum_{\substack{(y,v)\in E, \ v\in V}}\hspace{-15pt}e^c_{y,v,x,y,u,u'} = 0 \nonumber\\& \hspace{-33pt}c\in C, x\in V, y \in V, x\neq y, (u,u')\in G^c \label{eq:destination_2}
\end{align}
\begin{align} \label{eq:intermediate_1}
&\hspace{5pt}\sum_{\substack{(v,w)\in E, \ v\in V}} \hspace{-15pt}e^c_{v,w,x,y,u,u'} = \hspace{-20pt}\sum_{\substack{(w,v')\in E, \ v'\in V}}\hspace{-18pt}e^c_{w,v',x,y,u,u'} \nonumber\\ &\hspace{27pt} c\in C, w \in V, x\in V, y \in V, x \neq w, y \neq w, (u,u')\in G^c
\end{align}
\begin{align}
&\hspace{5pt}\sum_{\substack{(v,w)\in E, \ v\in V}}\hspace{-15pt}e^c_{v,w,x,y,u,u'} \le 1 \nonumber\\&\hspace{27pt} c\in C, w \in V, x\in V, y \in V, x \neq w, y\neq w,(u,u')\in G^c \label{eq:intermediate_2}
\end{align}
\begin{align} \label{eq:no_loop_1}
& \hspace{12pt}e^c_{v,v,x,y,u,u'} = 0 \nonumber\\ & \hspace{12pt} c\in C, v \in V, x \in V, y \in V, x \neq y, (u,u') \in G^c
\end{align}
\begin{align}
& \hspace{12pt}e^c_{v,v',x,x,u,u'} = 0 \nonumber\\ & \hspace{12pt} c\in C, x \in V, (u,u') \in G^c, (v,v') \in E,  v\neq v' \label{eq:no_loop_2}
\end{align}
Eq. \ref{eq:e_m_relationship} guarantees that the physical link $(v,v')$ can belong to a path between two nodes $x$ and $y$ for a virtual link $(u,u')$ of the SFC $c$ only if the two consecutive VNF requests or start/end points $u$ and $u'$ are mapped to nodes $x$ and $y$, respectively. The product $m^c_{u,x} \ m^c_{u',y}$ of binary variables must be linearized.
Eq. \ref{eq:source} and eq. \ref{eq:destination} are respectively the \emph{source} and \emph{destination} constraints. In fact, eq. \ref{eq:source} assures that the virtual link $(u,u')$ between two consecutive VNF requests or start/end points $u$ and $u'$ originates from one of the links connected to the node $x$, where the VNF request or start point $u$ is mapped to (eq. \ref{eq:source}), and it ends in one of the links of the node $y$, where the VNF request or end point $u'$ is mapped to (eq. \ref{eq:destination}). The products $e^c_{x,v,x,y,u,u'} \ m^c_{u,x} \  m^c_{u',y}$ and $e^c_{v,y,x,y,u,u'} \ m^c_{u,x} \ m^c_{u',y}$ must be linearized. In addition to source and destination constraints, the model also has to guarantee that no spurious links are selected. To do so, we have introduced eq. \ref{eq:source_2} and eq. \ref{eq:destination_2}. While mapping the virtual link $(u,u')$ for the SFC $c$ on a physical path between the nodes $x$ and $y$ where the VNF requests or start/end points $u$ and $u'$ are mapped to, no incoming physical link for the node $x$ (eq. \ref{eq:source_2}) and no outgoing link for the node $y$ (eq. \ref{eq:destination_2}) must be considered. Eq. \ref{eq:intermediate_1} and eq. \ref{eq:intermediate_2} are then the \emph{transit} constraints. While considering a generic node $w$ (neither source node $x$ nor destination node $y$ of a virtual link $(u,u')$), if one of its incoming links belongs to the path between nodes $x$ and $y$, then also one of its outgoing links must belong to the path (eq. \ref{eq:intermediate_1}). Without eq. \ref{eq:intermediate_2} multiple incoming/outgoing links could be considered, but in this paper we deal with unsplittable flows. Finally, eq. \ref{eq:no_loop_1} and eq. \ref{eq:no_loop_2} ensure a correct usage of self-loops, as introduced in Section \ref{sec:system_model}. A self-loop for an NFV node $v$ is used when two consecutive VNF requests or start/end points $u$ and $u'$ are mapped to the same node $x$, and cannot be used otherwise (eq. \ref{eq:no_loop_1}). Moreover, no physical link $(v,v')$ other than the self-loop is used when VNF requests or start/end points $u$ and $u'$ are mapped to the same node $x$ (eq. \ref{eq:no_loop_2}).

\subsubsection{Performance constraints}
\begin{align} \label{eq:bandwidth}
\hspace{-38pt}\sum_{\substack{c\in C, \ x,y\in V \\(u,u')\in G^c}}\hspace{-10pt}e^c_{v,v',x,y,u,u'} \ \delta^c_{u,u'} \le \beta_{v,v'} \quad (v,v')\in E
\end{align}
\begin{align} \label{eq:latency_costs}
 \hspace{15pt}\sigma^c &=\hspace{-10pt}\sum_{\substack{f\in F, \ v\in V \\ u\in U^c : \tau^c_u=f}}\hspace{-10pt} m^c_{u,v} \ (\text{CSW}_{\lat}(v)+\text{UP}_{\lat}(f,v)) \nonumber\\
&=\hspace{-10pt}\sum_{\substack{f\in F, \ v\in V \\ u\in U^c : \tau^c_u=f}}\hspace{-10pt} m^c_{u,v} \ (\sum_{\substack{f\in F}} \lceil c_{f,v} \rceil \omega_v + \lceil c_{f,v}\rceil \kappa_v)  \quad c\in C
\end{align}
\begin{align} \label{eq:latency}
\hspace{-45pt}\sum_{\substack{(v,v')\in E, \ x,y \in V \\ (u,u')\in G^c}} \hspace{-20pt}e^c_{v,v',x,y,u,u'} \ l_{v,v'} + \sigma^c \le \varphi^c \quad c\in C
\end{align}
\begin{align} \label{eq:user_indicator_1}
\hspace{-85pt}\sum_{f \in F} i_{f,v} \le \mathcal{M}_{F} \ a_v \quad v\in V
\end{align}
\begin{align}
 \hspace{-92pt}a_v \le \sum_{f \in F} i_{f,v} \quad v \in V  \label{eq:user_indicator_2}
\end{align}
Eq. \ref{eq:bandwidth} is the \emph{bandwidth constraint}. It assures that the overall bandwidth $\delta^c_{u,u'}$ requested by the virtual link $(u,u')$ of every SFC $c$ and mapped to the physical link $(v,v')$ cannot exceed the capacity of the link $\beta_{v,v'}$. 
In eq. \ref{eq:latency_costs} we compute $\sigma^c$, i.e., the overall context switching and upscaling latency costs per SFC $c$ as defined in Section \ref{sec:system_model}. Note that $\sigma^c$ refers to the latency introduced by \emph{all} the NFV nodes crossed by the SFC $c$. Eq. \ref{eq:latency} is then the \emph{latency constraint}. It assures that the latency introduced by the network between start and end points of a SFC $c$ cannot overcome the maximum tolerated latency $\varphi^c$. It takes into account both latency introduced by the propagation over physical links and by the NFV nodes due to upscaling and context switching (i.e., $\sigma^c$).
Note that eq. \ref{eq:latency_costs} requires the linearization of the product between the binary variable $m^c_{u,v}$ and the real variable $\sum_{\substack{f\in F}} \lceil c_{f,v} \rceil \omega_v + \lceil c_{f,v}\rceil \kappa_v$, as well as of the ceiling function $\lceil c_{f,v}\rceil$.
Finally, eq. \ref{eq:user_indicator_1} and eq. \ref{eq:user_indicator_2} assure that a node is marked as \emph{active} (i.e., $a_v=1$) only if at least one instance of any VNF $f$ is hosted by it. The big-M parameter $\mathcal{M}_F$ must be chosen such that $\mathcal{M}_F > |F|$.

\subsection{Heuristic Cost-aware Algorithm description} \label{sec:heuristic}
\algrenewcommand{\algorithmiccomment}[1]{ \textbackslash* \emph{#1} *\textbackslash}

\begin{algorithm}[t]
\caption{Heuristic Cost-aware Algorithm (HCA)} \label{alg:hca}
\begin{algorithmic}[1]
\State Sort SFCs by increasing latency \label{alg:sort}
\Repeat{}
\State Pick next SFC \label{alg:nextsc}
\Statex \hspace{12pt} \textbackslash* Start of \emph{Phase 1} *\textbackslash
\Repeat{}
\State Pick next VNF request in SFC \label{alg:nextvnf}
\If{$\exists$ instances of VNF in the network} \label{alg:instanceplaced}
\State Sort VNF instances by increasing \label{alg:sortvnf}
\Statex \hspace{45pt}distance from the last placed VNF instance
\Statex \hspace{45pt}or start point (\emph{current node})
\State Try to scale up VNF instances until success \label{alg:scaleup}
\Statex \hspace{45pt}or all VNF instances have been tried
\If{success} \label{alg:successvnf}
\State \textbf{continue}
\EndIf
\EndIf
\Statex \hspace{27pt}\Comment{Failure}
\State Sort NFV nodes by increasing residual capacity \label{alg:sortnodes}
\State Try placing new VNF instance on an NFV node \label{alg:selectnode}
\Statex \hspace{30pt}or all NFV nodes have been tried
\If{failed} \label{alg:failnode1}
\State \textbf{return}(infeasible) \label{alg:failnode2}
\EndIf
\Until{all VNF requests are chained} \label{alg:untilvnf}
\Statex \hspace{12pt}  \textbackslash* Start of \emph{Phase 2} *\textbackslash
\State Check end-to-end latency of embedded SFC against \label{alg:evlatency1}
\Statex \hspace{15pt}latency requirement
\If{success} \label{alg:successlatency1}
\State \textbf{continue} \label{alg:successlatency2}
\EndIf
\Statex \hspace{12pt}\Comment{Failure}
\State Release resources allocated in \emph{Phase 1} \label{alg:release}
\State Place chained VNF instances along the end-to-end  \label{alg:placeonsp}
\Statex \hspace{12pt} latency shortest path
\State Check end-to-end latency of embedded SFC against \label{alg:evlatency2}
\Statex \hspace{15pt}latency requirement
\If{failed} \label{alg:failedscembed}
\State \textbf{return}(infeasible) \label{alg:scinfeasible}
\EndIf
\Until{all SFCs are embedded} \label{alg:untilsc}
\State \textbf{return}(success) \label{alg:successalg}
\end{algorithmic}
\end{algorithm}
Our Heuristic Cost-aware Algorithm (HCA) sequentially embeds SFCs $c\in C$. The algorithm starts by assuming that the physical links always have enough bandwidth capacity to accommodate the bandwidth requested by each SFC. This means that the link  capacity is never a bottleneck, meaning that the network is overprovisioned as usually occurs in the core network segment.
The pseudocode of HCA is shown in Alg. \ref{alg:hca}. The main idea is to build, using a greedy procedure, an embedding solution for each SFC $c$ while trying to re-use already-placed VNF instances or already-active NFV nodes first (\emph{Phase 1}). Only if the latency requirement $\varphi^c$ for the SFC $c$ is not met after \emph{Phase 1}, a \emph{Phase 2} is started to improve the solution.

Before starting the embedding, SFCs are sorted in an increasing order according to $\varphi^c$  (line \ref{alg:sort}). This way, the most latency-sensitive SFCs are placed first.
Then, the next SFC to embed is selected (line \ref{alg:nextsc}) and \emph{Phase 1} starts.
As first step of \emph{Phase 1}, the next VNF request to chain for the SFC is selected (line \ref{alg:nextvnf}) and an already-placed VNF instance for that VNF request is searched in the network (line \ref{alg:instanceplaced}). If more than one VNF instances are available, they are sorted by increasing distance (line \ref{alg:sortvnf}), i.e., the one mapped to the closest NFV node $v$ from the \emph{current node}, according to the latency shortest paths (SPs), is selected first. The \emph{current node} can be either the start point, if the embedding has just started, or the NFV node where the last-selected VNF request has been mapped to. HCA then tries to scale up the processing resources $c_{f,v}$ of the selected VNF instance on the NFV node $v$ by a value $\pi_f$, i.e., $c_{f,v} \rightarrow c_{f,v}+\pi_f$ (line \ref{alg:scaleup}), in order to be able to process the additional aggregated traffic passing through the VNF request. This operation leads to an increase in both context switching and upscaling costs for the node $v$ since they both depend on $c_{f,v}$. If the increase of context switching/upscaling latency costs compromises the end-to-end latency of already-embedded SFCs having one or more VNF requests mapped to $v$ (i.e., for some already-placed SFC it happens that the updated end-to-end latency overcomes $\varphi^c$), the scale up fails and the next VNF instance, on another NFV node $v$, is checked (line \ref{alg:scaleup}). Otherwise, if successful, the scale up is executed, the VNF request is mapped to the VNF instance on $v$ and the latency SP between the \emph{current node} and the selected NFV node $v$ is used to steer the SFC aggregated traffic between such physical nodes. The scale up can also fail if it triggers an increase in the context switching/upscaling processing costs such that not enough residual processing capacity is available after the scale up itself. If the scale up fails for all the already-placed VNF instances (i.e., the check of line \ref{alg:successvnf} fails), the algorithm sorts the remaining NFV nodes by increasing residual capacity (line \ref{alg:sortnodes}) and tries to place a new VNF instance of size $\pi_f$ on an NFV node (line \ref{alg:selectnode}). This way, the most used NFV node $v$ is selected first. This operation leads to an increase in the context switching/upscaling costs, and also in this case the algorithm checks that \emph{(i)} the end-to-end latency for the already-embedded SFCs with one or more VNF requests mapped to $v$ is not compromised and \emph{(ii)} the residual processing capacity of the NFV node $v$, after updating the context switching/upscaling processing costs, is enough to host the new VNF instance. If such operation succeeds, the VNF request is mapped to the new VNF instance. The latency SP in the physical network between the \emph{current node} and the selected NFV node $v$ is then used to steer SFC aggregated traffic between such physical nodes. Note that the operation described in line \ref{alg:selectnode} can imply the activation of an NFV node, if not enough residual processing capacity is available on already-active NFV nodes.  If during \emph{Phase 1} the placement of a VNF request fails, it means that the related SFC cannot be embedded for lack of processing resources and the overall embedding process is aborted (lines \ref{alg:failnode1}-\ref{alg:failnode2}).
Conversely, if all VNF requests are greedily mapped to NFV nodes (line \ref{alg:untilvnf}), the NFV node where the last VNF request has been mapped is connected to the end node through the latency SP on the physical network. \emph{Phase 1} is then completed and an embedding solution for the SFC is found.

The greedy SFC embedding solution built according to the steps shown in \emph{Phase 1} tries to first use NFV nodes that are already active: this feature fits well with the objective of maximum consolidation. However, such solution can lead to a high end-to-end latency for the selected SFC, since the placement is unaware of the relative position of SFC start/end points with respect to NFV nodes where the chained VNF requests are placed. In fact, \emph{Phase 1} only assures that traffic among VNFs is locally steered according to latency SPs among those NFV nodes where VNFs are mapped, but there is no guarantee that the solution is effective in terms of end-to-end latency.

For this reason, a \emph{Phase 2} needs to be executed. \emph{Phase 2} starts with the evaluation of the end-to-end latency of the solution built in \emph{Phase 1} (line \ref{alg:evlatency1}). If such latency is less or equal to $\varphi^c$, the embedding for the selected SFC is successful (lines \ref{alg:successlatency1}-\ref{alg:successlatency2}): this happens typically for SFCs with a loose latency requirement. Otherwise, the solution is discarded: this means that the processing resources allocated in \emph{Phase 1} are released (line \ref{alg:release}) and involved VNF instances are scaled down ($c_{f,v} \rightarrow c_{f,v}-\pi_f$). A new embedding solution is then built based on the latency SP between the start and end point of the selected SFC (line \ref{alg:placeonsp}). This helps to minimize the end-to-end latency introduced by the links. In order to maximally consolidate the VNF instances, we activate an inactive NFV node on such SP and we place a new VNF instance $f$ of size $\pi_f$ for each chained VNF request on it, by also updating the context switching and upscaling costs for the NFV node. For the placement of those multiple VNFs, the algorithm chooses the NFV node with maximum processing capacity $\gamma_v$, to facilitate re-usability of the node for future embeddings. Finally, the end-to-end latency is checked again (line \ref{alg:evlatency2}). If the latency requirement $\varphi^c$ still cannot be met or no NFV node can be activated, there is no feasible solution (lines \ref{alg:failedscembed}-\ref{alg:scinfeasible}), otherwise the SFC embedding is completed and a new SFC is selected, until all SFCs have been successfully embedded (lines \ref{alg:untilsc}-\ref{alg:successalg}) or the operation fails during the embedding of other SFCs. Note that it is not always guaranteed that between start/end points of a SFC an NFV node exists on the latency SP. To overcome this issue, in \emph{Phase 2} the algorithm computes the latency $k$ shortest paths ($k$-SP), where $k$ is chosen to ensure that at least one path crossing at least one NFV node is explored.

\begin{figure}[t]
	\centering
	\includegraphics[width=0.8\columnwidth]{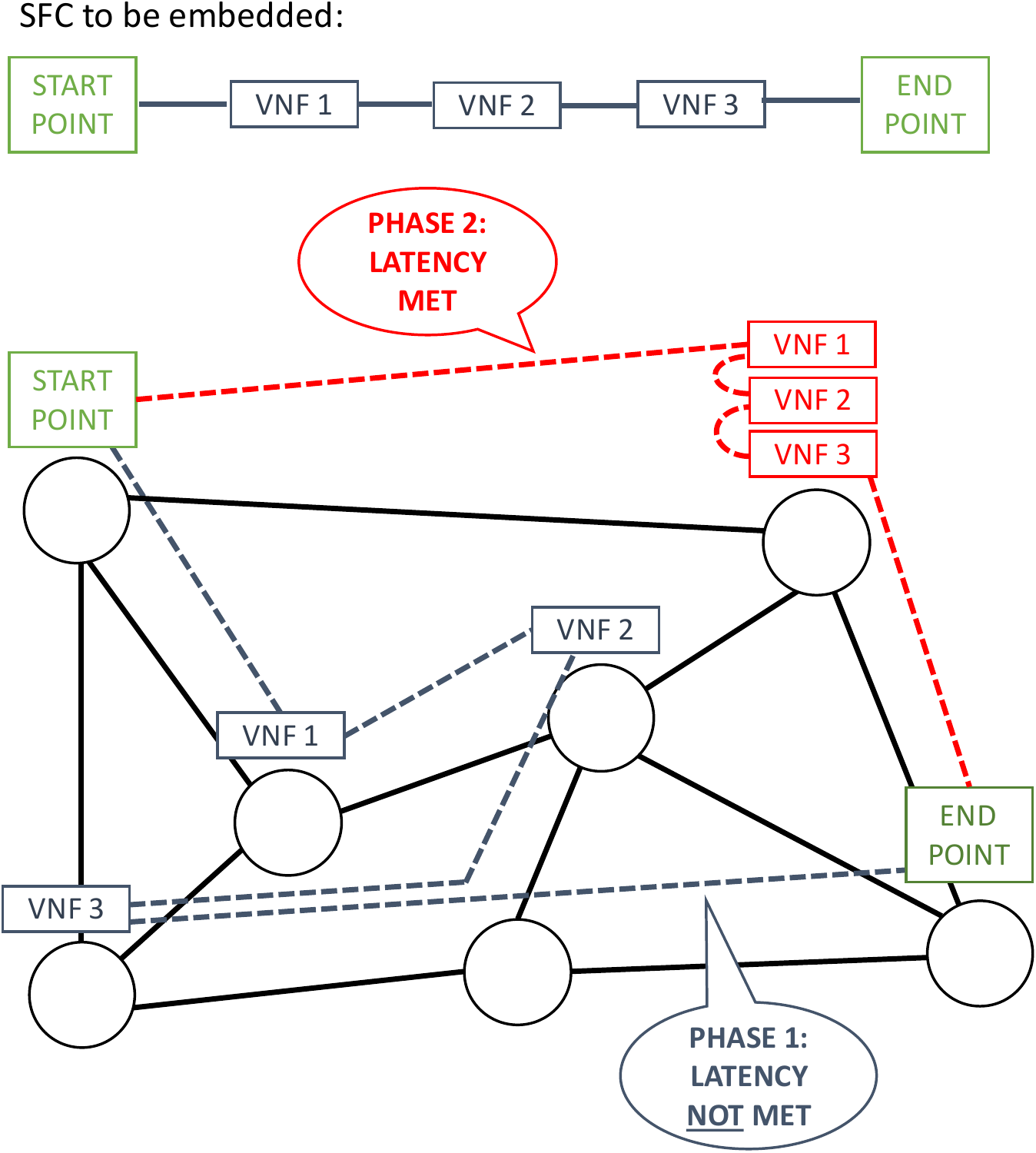}\caption{Pictorial example of HCA \emph{Phase 1} and \emph{Phase 2} execution \label{fig:hca_example}}
\end{figure}
Figure \ref{fig:hca_example} shows a pictorial example where a SFC concatenating three different VNF requests is embedded in the physical network according to \emph{Phase 1} steps, and then the solution is improved by \emph{Phase 2}.
It should also now be clear why as first step HCA sorts SFCs in an increasing order according to their latency requirement: the idea is to first activate all the NFV nodes thar are needed to meet the latency requirement for latency-sensitive SFCs, and then to reuse such NFV nodes to embed SFCs requiring looser latency, which do not need to have a latency-optimized end-to-end path.

\section{Computational results} \label{sec:results}
In this section we apply our algorithms to an example network to study how they can be used for VNF consolidation. We also show how the two processing-resource sharing cost parameters introduced in the paper impact on the cost for NFV implementation in different scenarios.

We first describe the computational settings, then we compare results by solving the ILP model and running HCA on some small-scale instances. Finally, we deepen our investigation by running our HCA on much larger-scale instances, and we compare our strategy with a state-of-the-art approach.

\begin{table}
	\caption{Processing Requirement (per user) for the VNFs \label{tab:vnf_processing} }
	\centering
	\begin{tabular}{|c|c|}
		\hline
		\centering Virtual Network Function & $\pi^{\user}$\\
		\hline
		\centering Network Address Translator (NAT) & 0.00092\\
		\centering Firewall (FW) & 0.0009\\
		\centering Traffic Monitor (TM) & 0.0133\\
		\centering WAN Optimization Controller (WOC) & 0.0054\\
		\centering Intrusion Detection Prevention System (IDPS) & 0.0107\\
		\centering Video Optimization Controller (VOC) & 0.0054\\
		\hline
	\end{tabular}
\end{table}
\begin{table} [t]
\caption{Performance Requirements for the SFCs \label{tab:sc_example} }
 	\centering
 \begin{tabular}{|c|c|c|c|}
 \hline
 \centering SFC & Chained VNFs & Latency & Bandwidth\\
 \centering  &  & $\varphi$& $\delta^{\user}$\\
 \hline
 \centering Web Service & NAT-FW-TM-WOC-IDPS & 500 ms & 100 kb/s\\
 \centering VoIP & NAT-FW-TM-FW-NAT & 100 ms & 64 kb/s\\
 \centering Video Streaming & NAT-FW-TM-VOC-IDPS & 100 ms & 4 Mb/s\\
 \centering Cloud Gaming & NAT-FW-VOC-WOC-IDPS & 60 ms & 4 Mb/s\\
 \hline
 \end{tabular}
 \end{table}
\subsection{Computational settings}
\begin{table*} [t]
\caption{Performance comparison between ILP and HCA \label{tab:comparison_results} }
 	\centering
 \begin{tabular}{|c|c|c|c|c||c|c|}
 \hline
 \centering & \multicolumn{2}{|c|}{$|C| = 3$, $N^{\user} = 300$, $N^{\iter}=100$} & \multicolumn{2}{|c||}{$|C| = 6$, $N^{\user} = 150$, $N^{\iter}=100$} & \multicolumn{2}{|c|}{$|C| = 8$, $N^{\user}=450$, $N^{\iter}=10$}\\
 \hline
 Latency Costs \cite{cerrato_vnf_2}& HCA & ILP & HCA & ILP & HCA & ILP\\
 \hline
 \hline
  \multicolumn{7}{|c|}{\textbf{Average number of active NFV nodes}}\\
 \hline
 \centering $\omega=0$ ms, $\kappa = 0$ ms &  $2.91 \pm 0.057$ & $2.91 \pm 0.057$	& $2.95 \pm 0.043$ & $2.95 \pm 0.043$ & $6.00 \pm 0.000$ & $6.00 \pm 0.000$ \\

  \centering $\omega = 0$ ms, $\kappa = 1.75$ ms & $2.95 \pm 0,052$ & $2.93\pm 0.058$ &	$2.99 \pm 0.034$ &	$2.97 \pm 0.034$ &	$6.11 \pm 0.260$   & $6.00 \pm 0.000$ \\

 \centering $\omega = 0.4$ ms, $\kappa = 0$ ms & $3.09 \pm 0.090$	& $3.07 \pm 0.086$ & $3.00 \pm 0.028$ & $2.99 \pm 0.020$ &	$7.86 \pm 0.640$ & $6.45 \pm 0.410$\\
 \hline
  \multicolumn{7}{l}{\bigstrut{\footnotesize Results are reported along with 95\% confidence intervals}} \\
  \hline
   \multicolumn{7}{|c|}{ \textbf{\% Infeasible computational instances }}\\
   \hline
  \centering $\omega=0$ ms, $\kappa = 0$ ms& 0 & 0 & 0 & 0 & 0 & 0 \\

  $\omega = 0$ ms, $\kappa = 1.75$ ms & 0 & 0 & 0 & 0 & 10 & 10\\

  $\omega = 0.4$ ms, $\kappa = 0$ ms & 0 & 0 &	0 &	0 &	30 & 10 \\
  \hline
  \hline
  \multicolumn{7}{|c|}{\textbf{Execution time per computational instance}}\\
  \hline
  $\omega=0$ ms, $\kappa = 0$ ms & 32.81 ms	 &	2.91 min   & 49.34 ms	 & 4.32 min	   &	2.01 s & 1.80 h\\

  $\omega = 0$ ms, $\kappa = 1.75$ ms & 33.27 ms	 &	3.23 min   & 49.90 ms	 & 4.87 min	   &	2.76 s & 13.70 h\\

  $\omega = 0.4$ ms, $\kappa = 0$ ms & 33.13 ms	 &	3.55 min   & 49.69 ms	 & 5.19 min	   &	2.35 s & 16.85 h\\
  \hline
 \end{tabular}
 \end{table*}

We consider the ISP physical network topology shown in Fig. \ref{fig:topology} with ten physical nodes ($|V|=10$) and fifteen physical links ($|E|=15$). This network topology is taken from the US Internet2 network \cite{Internet2}, considering only the nodes with advanced layer 3 services. The latency introduced by the physical links due to propagation and forwarding is in the order of milliseconds (the shortest link has $\lambda_{v,v'}=3$ ms, while the longest link has $\lambda_{v,v'}=13.5$ ms). We also assume that the bandwidth on the physical links cannot be a bottleneck ($\beta_{v,v'}=\infty \ \forall (v,v')\in E$) and that all the physical nodes are NFV nodes and equipped with a multi-core CPU ($\gamma_v=16 \ \forall v\in V$). Moreover, all the NFV nodes are always characterized by the same context switching and upscaling parameters, both in terms of latency ($\omega_v$ and $\kappa_v$) and processing ($\xi_v$ and $\mu_v$). We assume that latency and processing parameters (both concerning context switching and upscaling) are linearly dependent, according to another parameter $h$. This means that it always holds $\omega_v = h\xi_v$ and $\kappa_v = h \mu_v$. In our computational tests, unless otherwise specified, we set $h=0.01$.

We consider six different VNFs, reported in Table \ref{tab:vnf_processing}. Each VNF has a different processing requirement per user $\pi^{\user}$. The processing requirement for VNFs has been obtained according to middleboxes datasheets (see e.g. \cite{dellsonicwall}) as ratio between the number of adopted CPU cores and of flows supported by the middlebox. Even though this is just a possible and rough estimation for the processing requirement (see \cite{doriguzzi19} for a different strategy based on CPU cycles/s), it allows to understand which are the most processing-hungry VNFs: for example, according to our estimation, a Traffic Monitor is about 15 times more processing-hungry than a Firewall.
The six VNFs can be chained in different ways to provide four heterogeneous SFCs, reported in Table \ref{tab:sc_example}. Such SFCs represent four different services, i.e., Web Service (WS), VoIP, Video Streaming (VS) and Cloud Gaming (CG). Table \ref{tab:sc_example} shows also the performance requirements in terms of bandwidth $\delta$ and maximum tolerated latency $\varphi$ for each SFC, assuming that every virtual link $(u,u')$ of a SFC
requires the same bandwidth. Performance requirements for WS, VoIP and VS are well known, while performance requirements for CG, which is a more recent service gaining more and more interest by the research community due to the technical challenges it poses \cite{valls14}, are taken by \cite{shea13}.

 We show our results in different scenarios: a \emph{mixed scenario} and some \emph{homogeneous scenarios}. In the former we run a number $N^{\iter}$ of computational instances while uniformly randomizing the choice of SFC type to embed, among the four different types of SFCs, and of start/end points, among all the 10 physical nodes of the network. In the latter we only uniformly randomize the choice of start/end points, while we assume that only one type of SFC is embedded in the network.
 
 Note that Table \ref{tab:tunable_parameters} reports a summary of the tunable parameters in our evaluations to help the reader better understand the following results.
 
\begin{table}
	\caption{Tunable parameters in evaluations \label{tab:tunable_parameters} }
	\centering
	\begin{tabular}{|c|p{6cm}|}
		\hline
		\centering Tunable param. & Description\\
		\hline
		\centering $h$ & Dependency between latency/processing penalties\\
		\centering $\omega$ & Context switching latency\\
		\centering $\kappa$ & Upscaling latency\\
		\centering  $N^{\iter}$ & Number of computational instances\\
		\centering  $|C|$ & Number of SFCs to be embedded\\
		\centering $N^{\user}$ & Number of users per SFC\\
		
		\hline
	\end{tabular}
\end{table}

\subsection{ILP and HCA performance comparison}
We solved the ILP model using ILOG CPLEX solver, while we implemented HCA in Matlab. All the computational tests were run on a workstation equipped with $8 \times 2$ GHz CPU cores and with 32 GB of RAM. To compare results obtained by ILP and HCA we have focused on three different processing-resource sharing cost settings:
\begin{itemize}
\item  A \emph{No costs} setting ($\omega = 0$ ms, $\kappa = 0$ ms);
\item An \emph{Only upscaling costs} setting ($\omega = 0$ ms, $\kappa = 1.75$ ms);
\item An \emph{Only context switching costs} setting ($\omega = 0.4$ ms, $\kappa = 0$ ms).
\end{itemize}

We chose aforementioned values for $\omega$ and $\kappa$ since they lead to NFV node latencies of few milliseconds, as in \cite{cerrato_vnf_1}. 

In this evaluation we consider a \emph{mixed scenario}. For any of the cost settings, we have simulated an increasing number of SFCs, i.e.,  $|C|= 3$, $6$, and $8$. For $|C|=3$ and $|C|=6$ the overall number of users in the network is the same ($|C|N^{\user} = 900$ in both cases), while for $|C|=8$ a higher number of users  ($|C|N^{\user}=3600$) is considered, leading thus to a more loaded network scenario. We consider a number of randomized instances $N^{\iter}=100$ for $|C|=3$ and $|C|=6$, while a number of randomized instances of $N^{\iter}=10$ for $|C|=8$. Results are shown in Table \ref{tab:comparison_results}. We report the \emph{average number of active NFV nodes} along with the $95\%$ confidence interval. We also report the \emph{percentage of infeasible computational instances} and the \emph{execution time per computational instance}. We can see how for $|C| = 3$ and $|C|=6$, HCA closely matches results obtained by solving the ILP for all the cost settings. There are not infeasible instances and HCA allows to obtain a near-optimal solution in a computational time in the order of milliseconds per instance, while some minutes are needed to solve the ILP model.
In case of $|C|=8$ and more loaded network the results are slightly different. First of all, for both HCA and ILP some instances are infeasible. This can happen because \emph{(i)} the latency requirement for some of the SFCs cannot be met or \emph{(ii)} there is not enough processing in the NFV nodes to place all the VNFs. Especially, for the \emph{Only context switching costs} there is a higher infeasibility percentage for HCA (30\%) than for the ILP (10\%). This means that part of the ILP feasible solutions cannot be explored by HCA. Additionally, in such case, HCA activates on average about 1.4 NFV nodes more than the optimal solution. However, execution times per computational instance are in the order of seconds for HCA, while the ILP needs several hours to be solved. Especially, the needed time to solve it for the \emph{Only context switching costs} (16.85 h) and \emph{Only upscaling costs} (13.70 h) settings is about 10 times higher than the \emph{No costs} setting solving time (1.80 h). This happens because in the \emph{Only context switching costs} and \emph{Only upscaling costs} settings the solver must compute $\psi_v \ge 0$ (eq. \ref{eq:processing_costs}) and $\sigma^c \ge 0$ (eq. \ref{eq:latency_costs}). Those terms include the ceiling function $\lceil c_{f,v} \rceil$, which is nonlinear and must be linearized. This is not true for the \emph{No costs} setting, since $\omega=\kappa=0$ implies $\psi_v=\sigma^c=0$ and thus $\lceil c_{f,v} \rceil$ must not be computed by the solver, leading to a simpler ILP model to solve.

\subsection{Mixed scenario}
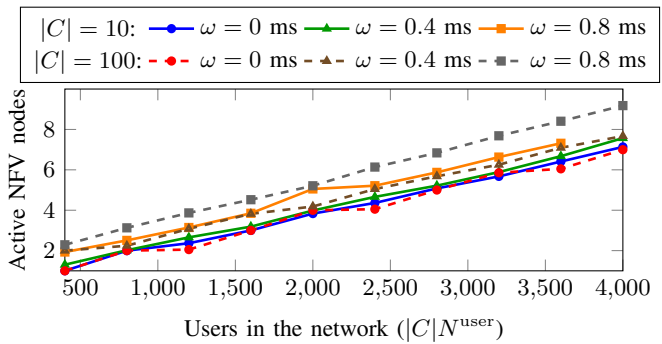
\begin{figure} [t]
     \centering
       \begin{tikzpicture}[font=\small]
         \begin{axis}[
     height=4 cm,
     width=9 cm,
     ylabel={Active NFV nodes},
     xlabel={Users in the network ($|C|N^{\user}$)},
     xmin=400,
     xmax=4000,
     ymin=1,
     ymax=10,
     xtick={500,1000,...,4000},
     ytick={2,4,...,8},
     legend style={at={(-17pt,100 pt)},anchor=north west},
     legend columns=4,
     ]
           \pgfplotstableread{csw_mix.dat}{\comp}
           \addlegendimage{empty legend}
           \addlegendentry{$|C|=10$:}
           \addplot [color=blue,mark=*, mark size = 1.25pt, line width = 1] table [x index=0, y index=1] {\comp};
           \addlegendentry{$\omega=0$ ms}
           \addplot[color=green!60!black, mark=triangle,mark size = 1.25pt, line width = 1] table [x index=0, y index=2] {\comp};
           \addlegendentry{$\omega=0.4$ ms}
           \addplot[color=orange, mark=square*,mark size = 1.25pt, line width = 1] table [x index=0, y index=3] {\comp};
           \addlegendentry{$\omega=0.8$ ms}
           \addlegendimage{empty legend}
           \addlegendentry{$|C|=100$:}
           \addplot [color=red,mark=*, mark size = 1.25pt, line width = 1, mark options={solid},dashed] table [x index=0, y index=4] {\comp};
           \addlegendentry{$\omega=0$ ms}
           \addplot[color=brown!60!black,mark=triangle,mark size = 1.25pt, line width = 1, mark options={solid},dashed] table [x index=0, y index=5] {\comp};
           \addlegendentry{$\omega=0.4$ ms}
           \addplot[color=black!60!white,mark=square*,mark size = 1.25pt, line width = 1,mark options={solid},dashed] table [x index=0, y index=6] {\comp};
           \addlegendentry{$\omega=0.8$ ms}
          \end{axis}
       \end{tikzpicture}
       \caption{Number of active NFV nodes as a function of the overall number of users in the network, while considering the impact of different \emph{latency context switching costs} $\omega$  and different numbers of SFCs $|C|$ in the \emph{mixed scenario} ($\kappa=0$, $N^{\iter}=1000$)}
       \label{fig:csw_mix}
     \end{figure}

After having verified the effectiveness of HCA, we ran it for larger-scale instances to deepen the study. In this section we show the results obtained for the mixed scenario. Figure \ref{fig:csw_mix} shows the impact of different context switching costs, by varying $\omega$, on the number of active NFV nodes. This is evaluated as a function of the overall number of users in the network $|C|N^{\user}$. Upscaling costs are considered negligible ($\kappa=0$). We consider $N^{\iter}=1000$ and the embedding of $|C|=10$ and $|C|=100$ SFCs. The overall number of users in the network is equally split among the number of SFCs $|C|$ (i.e., if the users in the network are 1000 and $|C|=100$ then $N^{\user}=10$, while if $|C|=10$ then $N^{\user}=100$). This way, we can evaluate the impact of the number of SFCs to embed on VNF consolidation without changes in the overall network load. Figure \ref{fig:cup_mix} shows instead the impact of different upscaling costs, by varying $\kappa$, when context switching costs are negligible ($\omega=0$). We adopt the same settings as for Fig. \ref{fig:csw_mix}. In both cases we plot only values for which the percentage of infeasible instances is less than 20\%. 
\begin{figure} [t]
    \centering
      \begin{tikzpicture}[font=\small]
        \begin{axis}[
    height=4 cm,
    width=9 cm,
    ylabel={Active NFV nodes},
    xlabel={Users in the network ($|C|N^{\user}$)},
    xmin=400,
    xmax=4000,
    ymin=1,
    ymax=10,
    xtick={500,1000,...,4000},
    ytick={2,4,...,8},
    legend style={at={(-20pt,100 pt)},anchor=north west},
    legend columns=4,	
    ]
          \pgfplotstableread{cup_mix.dat}{\comp}
          \addlegendimage{empty legend}
          \addlegendentry{$|C|=10$:}
          \addplot [color=blue,mark=*, mark size = 1.25pt, line width = 1] table [x index=0, y index=1] {\comp};
          \addlegendentry{$\kappa=0$ ms}
          \addplot[color=green!60!black, mark=triangle,mark size = 1.25pt, line width = 1] table [x index=0, y index=2] {\comp};
          \addlegendentry{$\kappa=1.75$ ms}
          \addplot[color=orange, mark=square*,mark size = 1.25pt, line width = 1] table [x index=0, y index=3] {\comp};
          \addlegendentry{$\kappa=3.5$ ms}
          \addlegendimage{empty legend}
          \addlegendentry{$|C|=100$:}
          \addplot [color=red,mark=*, mark size = 1.25pt, line width = 1, mark options={solid},dashed] table [x index=0, y index=4] {\comp};
          \addlegendentry{$\kappa=0$ ms}
          \addplot[color=brown!60!black,mark=triangle,mark size = 1.25pt, line width = 1, mark options={solid},dashed] table [x index=0, y index=5] {\comp};
          \addlegendentry{$\kappa=1.75$ ms}
          \addplot[color=black!60!white,mark=square*,mark size = 1.25pt, line width = 1,mark options={solid},dashed] table [x index=0, y index=6] {\comp};
          \addlegendentry{$\kappa=3.5$ ms}
         \end{axis}
      \end{tikzpicture}
      \caption{Number of active NFV nodes as a function of the overall number of users in the network, while considering the impact of different \emph{latency upscaling costs} $\kappa$ and different numbers of SFCs $|C|$ in the \emph{mixed scenario} ($\omega=0$, $N^{\iter}=1000$)}
      \label{fig:cup_mix}
    \end{figure}
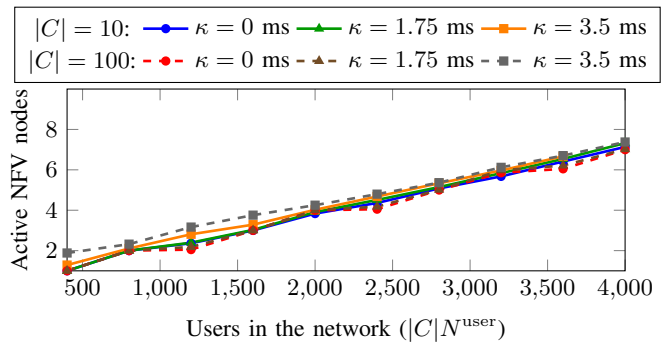
    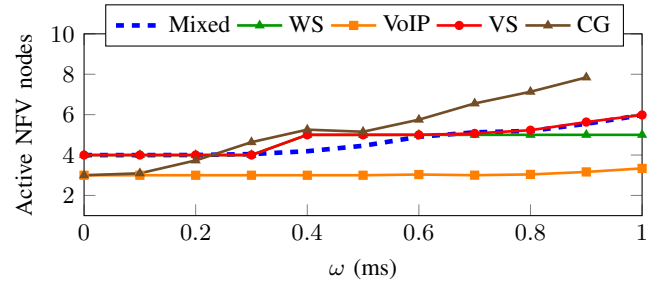
\begin{figure} [t]
    	\centering
    	\begin{tikzpicture}[font=\small]
    	\begin{axis}[
    	height=4 cm,
    	width=9 cm,
    	ylabel={Active NFV nodes},
    	xlabel={$\omega$ (ms)},
    	xmin=0,
    	xmax=1,
    	ymin=1,
    	ymax=10,
    	xtick={0,0.2,...,1},
    	ytick={2,4,...,10},
    	legend style={at={(8pt,80 pt)},anchor=north west},
    	legend columns=5,
    	]
    	\pgfplotstableread{csw_hom.dat}{\comp}
    	\addplot [color=blue, line width = 1.75, dashed] table [x index=0, y index=1] {\comp};
    	\addplot[color=green!60!black, mark=triangle,mark size = 1.25pt, line width = 1] table [x index=0, y index=2] {\comp};
    	\addplot[color=orange, mark=square*,mark size = 1.25pt, line width = 1] table [x index=0, y index=3] {\comp};
    	\addplot [color=red,mark=*, mark size = 1.25pt, line width = 1] table [x index=0, y index=4] {\comp};
    	\addplot[color=brown!60!black,mark=triangle,mark size = 1.25pt, line width = 1] table [x index=0, y index=5] {\comp};
    	\legend{Mixed, WS, VoIP, VS, CG}
    	\end{axis}
    	\end{tikzpicture}
    	\caption{Number of active NFV nodes as a function of the \emph{latency context switching costs} parameter $\omega$, considering four  \emph{homogeneous scenarios} according to the SFCs of Tab. \ref{tab:sc_example} ($|C|=100$, $N^{\user}=20$, $\kappa=0$,~$N^{\iter}=1000$)}
    	\label{fig:csw_hom}
    \end{figure}
    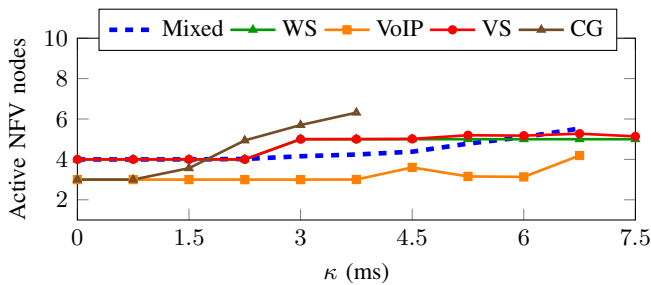
\begin{figure} [t]
    	\centering
    	\begin{tikzpicture}[font=\small]
    	\begin{axis}[
    	height=4 cm,
    	width=9 cm,
    	ylabel={Active NFV nodes},
    	xlabel={$\kappa$ (ms)},
    	xmin=0,
    	xmax=7.5,
    	ymin=1,
    	ymax=10,
    	xtick={0,1.5,...,7.5},
    	ytick={2,4,...,10},
    	legend style={at={(8pt,80 pt)},anchor=north west},
    	legend columns=5,
    	]
    	\pgfplotstableread{cup_hom.dat}{\comp}
    	\addplot [color=blue, line width = 1.75, dashed] table [x index=0, y index=1] {\comp};
    	\addplot[color=green!60!black, mark=triangle,mark size = 1.25pt, line width = 1] table [x index=0, y index=2] {\comp};
    	\addplot[color=orange, mark=square*,mark size = 1.25pt, line width = 1] table [x index=0, y index=3] {\comp};
    	\addplot [color=red,mark=*, mark size = 1.25pt, line width = 1] table [x index=0, y index=4] {\comp};
    	\addplot[color=brown!60!black,mark=triangle,mark size = 1.25pt, line width = 1] table [x index=0, y index=5] {\comp};
    	\legend{Mixed, WS, VoIP, VS, CG}
    	\end{axis}
    	\end{tikzpicture}
    	\caption{Number of active NFV nodes as a function of the \emph{latency upscaling costs} parameter $\kappa$, considering four  \emph{homogeneous scenarios} according to the SFCs of Tab. \ref{tab:sc_example} ($|C|=100$, $N^{\user}=20$, $\omega=0$, $N^{\iter}=1000$)}
    	\label{fig:cup_hom}
    \end{figure}
The curves for both Fig. \ref{fig:csw_mix} and Fig. \ref{fig:cup_mix} show a non-decreasing trend. In fact, increasing the number of users in the network implies an increase in the processing requirements $\pi$ for any placed VNF instance. Thus, more NFV nodes need to be activated to accommodate bigger VNFs. By comparing Fig. \ref{fig:csw_mix} and Fig. \ref{fig:cup_mix}, we can then notice two different trends. In Fig. \ref{fig:csw_mix}, by increasing the value of $\omega$, the relative difference between $|C|=10$ and $|C|=100$ curves strongly increases. This is not true for Fig. \ref{fig:cup_mix}, where such relative difference does not significantly vary. Moreover, in Fig. \ref{fig:cup_mix} all the curves show a weak dependence on the upscaling costs and on the number of SFCs, meaning that in general upscaling costs have less impact than context switching costs on the number of active NFV nodes.

\subsection{Homogeneous scenarios}
Here we consider results obtained in four different homogeneous scenarios. Figure \ref{fig:csw_hom} shows the number of active NFV nodes in the network as a function of the context switching costs parameter $\omega$ for the SFCs of Table \ref{tab:sc_example}. As we can see, SFCs with different requirements and chaining different VNFs have diverse impact on the cost for NFV implementation, measured by the number of active NFV nodes. The difference is mainly due to the distinct impact of both context switching and processing requirement of VNFs chained by SFCs. Looking at Tab. \ref{tab:vnf_processing} and Tab. \ref{tab:sc_example}, it is easy to see how WS and VS SFCs concatenate VNFs that, on average, have more processing requirement per user than VNFs chained by VoIP and Cloud Gaming SFCs. This explains why, for small values of $\omega$, the number of active NFV nodes is the same for WS/VS and for VoIP/CG homogeneous scenarios. Increasing the context switching parameter $\omega$, in general, leads to an increase in the number of active NFV nodes because both processing and latency context switching costs increase, making it harder to meet VNF processing and SFC latency requirements while activating a small number of nodes. However, the impact of context switching for VoIP homogeneous scenario starts to be noticeable only for high values of $\omega$, since the latency requirement is not very strict (100 ms) and the processing requirement for its chained VNFs is low. Additionally, the impact of context switching costs is very similar for WS and VS despite for high values of $\omega$. This happens because the average processing requirement for the VNFs chained by WS and VS SFCs, as recalled earlier, is very similar. However, for $\omega>0.7$ ms, the curves start diverging, being the number of active nodes for VS higher than for WS. In fact, VS SFCs have a stricter latency requirement than WS SFCs and, starting from $\omega>0.7$ ms, latency introduced by NFV nodes due to context switching becomes significant. This implies that more nodes need to be activated to meet the VF SFCs latency requirement. Finally, CG SFCs have a very strict latency requirement (60 ms). For this reason,  context switching costs have a very strong impact on the cost for NFV implementation. A high number of NFV nodes must be activated because, in order to guarantee the latency requirement, all VNFs must be placed on a end-to-end physical path between start/end points of each SFC close to the latency shortest path. It is also important to note how the curve of mixed scenario leads more or less to average values with respect to the curves representing homogeneous scenarios. For values of $\omega>0.4$ ms, latency introduced by NFV nodes due to context switching starts to affect the placement of VNFs, which must ensure that latency requirement for latency-sensitive SFCs (i.e., CG) in the mix is met. Similar considerations can be made for the upscaling costs, whose results are shown in Fig. \ref{fig:cup_hom}.

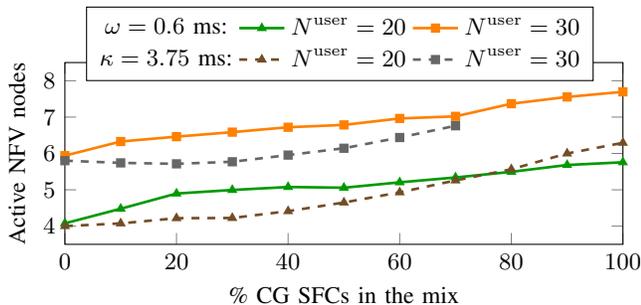
\begin{figure} [t]
	\centering
	\begin{tikzpicture}[font=\small]
	\begin{axis}[
	height=4 cm,
	width=9 cm,
	ylabel={Active NFV nodes},
	xlabel={\% CG SFCs in the mix},
	xmin=0,
	xmax=100,
	ymin=3.5,
	ymax=8.5,
	xtick={0,20,...,100},
	ytick={4,...,10},
	legend style={at={(8pt,90 pt)},anchor=north west},
	legend columns=3,	
	]
	\pgfplotstableread{cg.dat}{\comp}
	\addlegendimage{empty legend}
	\addlegendentry{$\omega=0.6$ ms:}
	\addplot [color=green!60!black,mark=triangle*, mark size = 1.25pt, line width = 1] table [x index=0, y index=2] {\comp};
	\addlegendentry{$N^{\user}=20$}
	\addplot[color=orange, mark=square*,mark size = 1.25pt, line width = 1] table [x index=0, y index=3] {\comp};
	\addlegendentry{$N^{\user}=30$}
	\addlegendimage{empty legend}
	\addlegendentry{$\kappa=3.75$ ms:}
	\addplot[color=brown!60!black,mark=triangle*,mark size = 1.25pt, line width = 1, mark options={solid},dashed] table [x index=0, y index=5] {\comp};
	\addlegendentry{$N^{\user}=20$}
	\addplot [color=black!60!white,mark=square*, mark size = 1.25pt, line width = 1, mark options={solid},dashed] table [x index=0, y index=6] {\comp};
	\addlegendentry{$N^{\user}=30$}
	\end{axis}
	\end{tikzpicture}
	\caption{Number of active NFV nodes as a function of the percentage of Cloud Gaming SFCs in the mix considering context switching latency costs $\omega=0.6$ ms, upscaling latency costs $\kappa=3.75$ ms and different number of aggregated users $N^{\user}$ per SFC ($|C|=100$, $N^{\iter}=1000$)}
	\label{fig:cg}
\end{figure}
Since the CG homogeneous scenario has the greatest impact on the cost for NFV implementation, it is interesting to investigate how different percentages of CG SFCs in the mix influence the number of NFV active nodes. Figure \ref{fig:cg} shows the number of active NFV nodes as a function of the percentage of CG SFCs in the mix. Shown results focus on a total number of SFCs $|C|=100$, consider different values of $N^{\user}$ ($N^{\user}=20$ and $N^{\user}=30$) and examine both context switching ($\omega=0.6$ ms) and upscaling ($\kappa=3.75$ ms). Results show that, concerning context switching costs, even a small number of CG SFCs in the mix has a strong impact on the number of active NFV nodes. For example, in case of $\omega=0.6$ ms and $N^{\user}=20$, having $20\%$ of CG SFCs in the mix is enough to significantly increase the average number of active NFV nodes with respect to the case where no CG SFCs must be embedded ($0\%$). This does not happen with upscaling costs. In this case, for $\kappa=3.75$ ms and $N^{\user}=20$, it is possible to notice a significant increase in the number of NFV active nodes only for percentages above 50\%. These results are in line with results shown previously for the mixed scenario, and we can then in general conclude that context switching costs have a much stronger impact on the cost for NFV implementation than upscaling costs.     
     \begin{figure} [t]
     	\centering
     	\begin{tikzpicture}[font=\small]
     	\begin{axis}[
     	height=4 cm,
     	width=9 cm,
     	ylabel={Active NFV nodes},
     	xlabel={Users in the network ($|C|N^{\user}$)},
     	xmin=400,
     	xmax=4000,
     	ymin=1,
     	ymax=10,
     	xtick={500,1000,...,4000},
     	ytick={2,4,...,8},
     	legend style={at={(-7pt,97 pt)},anchor=north west},
     	legend columns=3,
     	]
     	\pgfplotstableread{Comparison_SoTA_nodes.dat}{\comp}
     	\addlegendimage{empty legend}
     	\addlegendentry{$|C|=10$:}
     	\addplot [color=blue,mark=*, mark size = 1.25pt, line width = 1, dashed] table [x index=0, y index=1] {\comp};
     	\addlegendentry{SOTA}
     	\addplot[color=green!60!black, mark=triangle,mark size = 1.25pt, line width = 1] table [x index=0, y index=2] {\comp};
     	\addlegendentry{Processing-resource sharing}
     	\addlegendimage{empty legend}
     	\addlegendentry{$|C|=100$:}
     	\addplot [color=red,mark=*, mark size = 1.25pt, line width = 1, mark options={solid}, dashed] table [x index=0, y index=3] {\comp};
     	\addlegendentry{SOTA}
     	\addplot[color=brown!60!black,mark=triangle,mark size = 1.25pt, line width = 1, mark options={solid}] table [x index=0, y index=4] {\comp};
     	\addlegendentry{Processing-resource sharing}
     	\end{axis}
     	\end{tikzpicture}
     	\caption{Number of active NFV nodes as a function of the of the overall number of users in the network, while considering SOTA and our proposed \emph{node model} for different numbers of SFCs $|C|$ in the \emph{mixed scenario} ($h=0$, $N^{\iter}=1000$)}
     	\label{fig:comparison_nodes}
     \end{figure}
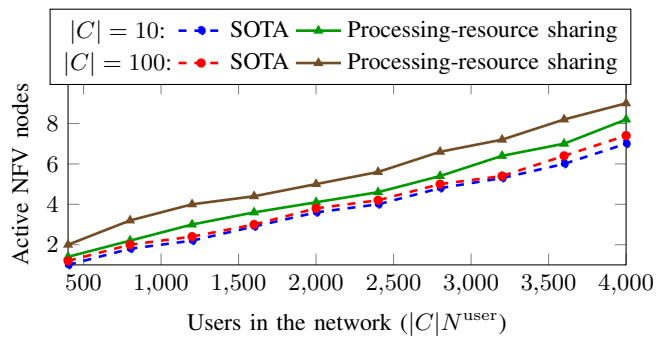
     
\subsection{Comparison with the state of the art}
Existing NFV node models for the estimation of expected introduced latency do not consider any processing-resource sharing aspect. The most common adopted model, which we call state-of-the-art (SOTA), merely considers a non-linear increase in the node latency as a function of the node utilization \cite{chua16}, neglecting any upscaling or context switching costs. In this subsection, we compare SFC embedding results obtained with our proposed processing-resource sharing node model, which relies on the costs introduced in Section \ref{sec:model_processing_resource_sharing}, with results obtained by adopting the SOTA model proposed in \cite{chua16}: to do so, we have modified HCA to embed SFCs according to such simplified model. Specifically, considering $P_v=\sum_{f\in F} \frac{c_{f,v}}{\gamma_v}$ as NFV node utilization, latency introduced by any NFV node $v$, for any VNF instance hosted by that node, is computed in the following way: 
\begin{equation}
\text{SOTA}_{\lat}(v) = \frac{P_v-[1+K(1-P_v)]P_v^{K+1}}{L(1-P_v)(1-P_v^K)}
\end{equation}
As per \cite{chua16}, we set $K=100$ and $L=10$. The computed node latencies are in the order of milliseconds, as the ones obtained with our model when $\omega=0.4$ ms and $\kappa=1.75$ ms, which are the values adopted in this subsection. Note also that, to simplify the evaluation, we assume no processing penalties, i.e., $h=0$.

\begin{figure} [t]
	\centering
	\begin{tikzpicture}[font=\small]
	\begin{axis}[
	height=4 cm,
	width=9 cm,
	ylabel={Average SFC latency},
	xlabel={Users in the network ($|C|N^{\user}$)},
	xmin=400,
	xmax=4000,
	ymin=0,
	ymax=100,
	xtick={500,1000,...,4000},
	ytick={20,40,...,80},
	legend style={at={(-7pt,92pt)},anchor=north west},
	legend columns=3,
	]
	\pgfplotstableread{Comparison_SoTA_latencies.dat}{\comp}
	\addlegendimage{empty legend}
	\addlegendentry{$|C|=10$:}
	\addplot [color=blue,mark=*, mark size = 1.25pt, line width = 1, dashed] table [x index=0, y index=1] {\comp};
	\addlegendentry{SOTA}
	\addplot[color=green!60!black, mark=triangle,mark size = 1.25pt, line width = 1] table [x index=0, y index=2] {\comp};
	\addlegendentry{Processing-resource sharing}
	\addlegendimage{empty legend}
	\addlegendentry{$|C|=100$:}
	\addplot [color=red,mark=*, mark size = 1.25pt, line width = 1, mark options={solid}, dashed] table [x index=0, y index=3] {\comp};
	\addlegendentry{SOTA}
	\addplot[color=brown!60!black,mark=triangle,mark size = 1.25pt, line width = 1, mark options={solid}] table [x index=0, y index=4] {\comp};
	\addlegendentry{Processing-resource sharing}
	\end{axis}
	\end{tikzpicture}
	\caption{Average SFC end-to-end latency as a function of the of the overall number of users in the network, while considering SOTA and our proposed \emph{node models} and different numbers of SFCs $|C|$ in the \emph{mixed scenario} ($h=0$, $N^{\iter}=1000$)}
	\label{fig:comparison_latencies}
\end{figure}
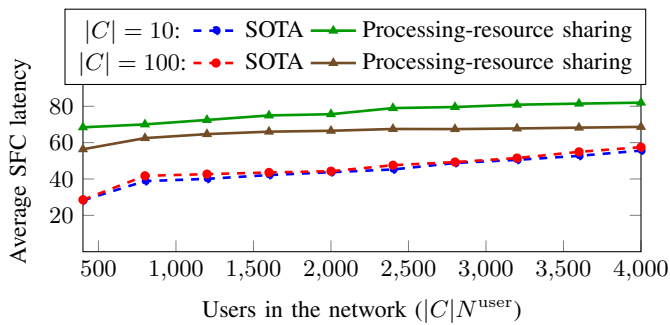

Since the SOTA model does not take into consideration how the processing capacity, in terms of number of cores, is assigned to different VNFs in an NFV node, it naturally misses to consider any arising processing-resource sharing latency penalty. For this reason, it tends to underestimate the NFV node latency, leading to greater VNF consolidations. This is shown in Figs. \ref{fig:comparison_nodes} and \ref{fig:comparison_latencies}.

Fig. \ref{fig:comparison_nodes} shows the impact of the two different node models on the number of active NFV nodes. This is investigated as a function of the overall number of users in the network $|C|N^{\user}$ in a mixed scenario. What can be seen is that the number of active NFV nodes by adopting the SOTA model is lower than in the case our processing-resource sharing model is considered, meaning that the SOTA model consolidates VNFs in less NFV nodes. However, being the curves for $|C|=10$ and $|C|=100$  overlapped for the SOTA model, it means that the number of active NFV nodes is invariant to $|C|$. Conversely, our model activates more NFV nodes as $|C|$ increases. This happens because with an increase of $|C|$ more VNF instances must be embedded in the network and more processing-resource sharing penalties arise. This is not true for the SOTA model, where the only relevant parameter is node utilization, which is roughly the same also in case of different $|C|$. Our model, thus, avoids an excessive VNF consolidation (especially when $|C|$ is larger) that could compromise the SFC end-to-end latency, especially for latency-sensitive SFCs.

Fig. \ref{fig:comparison_latencies} shows instead the average experienced end-to-end SFC latency, while considering the two models, in the same mixed scenario. As expected, the average latency is lower while adopting the SOTA model, and does not depend on $|C|$. This latency underestimation is mostly dangerous for latency-sensitive SFCs (e.g. CG SFCs), which would be embedded on paths that are estimated to meet their end-to-end latency requirements, but in reality may lead to much higher latency penalties due to processing-resource sharing.

\section{Conclusion} \label{sec:conclusion}
In this paper, we investigated the impact on the network cost of processing-resource sharing among VNFs in NFV, when multiple SFCs must be embedded in the network. VNF placement and distribution on NFV nodes lead to two different types of costs: upscaling and context switching costs, which affect the placement of VNFs and the embedding of SFCs. We first focused on the mathematical modeling of NFV nodes, VNFs, SFCs and of processing-resource sharing costs. Then, we formulated an ILP model and we designed a heuristic algorithm, called HCA, to evaluate the impact of such costs on VNF consolidation. We showed that HCA allows to obtain a near-optimal solution in a much shorter time than solving the ILP model. We then gathered some numerical results by focusing our attention on the placement of practical SFCs. Results showed that, in the considered ISP network, context switching costs have a greater impact on VNF consolidation than upscaling costs. Besides, such costs strongly affect NFV consolidation when SFCs with a very strict latency requirement, such as Cloud Gaming SFCs, must be embedded in the network. We also showed that this aspect cannot be captured by state-of-the-art node models neglecting processing-resource sharing aspects. 

Several issues remain open for future research. In fact, processing-resource sharing is just one of the possible resource sharing issues. Other resource sharing issues concerning memory and storage could be investigated. Moreover, VNF consolidation is not the only possible objective that a network operator is interested to achieve. Many other objectives, also taking into account bandwidth resources on the physical links, could be taken into consideration.

\bibliographystyle{IEEEtran}
\bibliography{bibliografia}
\vspace{-35pt}
\begin{IEEEbiography}[{\includegraphics[width=1in,height=1.25in,clip,keepaspectratio]{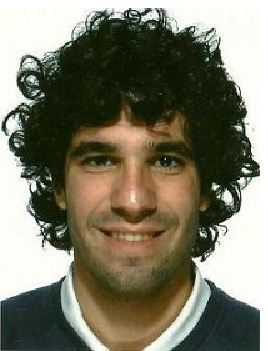}}]%
	{Marco Savi} is a postdoctoral researcher at FBK CREATE-NET, Trento, Italy. He received his PhD degree in Information Technology (Telecommunication engineering) in 2016 from Politecnico di Milano. His research interests mainly focus on the design and optimization of telecommunication networks, especially optical and 5G networks, and cloud computing. Dr. Savi has been involved in some European research projects advancing access and core network technologies.	
\end{IEEEbiography}
\vspace{-35pt}
\begin{IEEEbiography}[{\includegraphics[width=1in,height=1.25in,clip,keepaspectratio]{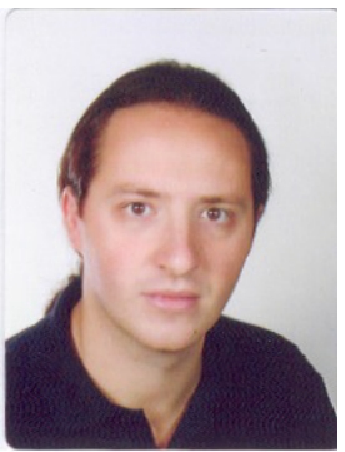}}]%
	{Massimo Tornatore} is an associate professor at Politecnico di Milano, Italy, and an adjunct associate professor at the University of California, Davis, USA.  He is author of about 200 technical papers (with 7 best paper awards) and his research interests include the design and engineering of telecom and cloud networks, through optimization and simulation. Prof. Tornatore received a PhD degree in 2006 from Politecnico di Milano. 	
\end{IEEEbiography}
\vspace{-35pt}
\begin{IEEEbiography}[{\includegraphics[width=1in,height=1.25in,clip,keepaspectratio]{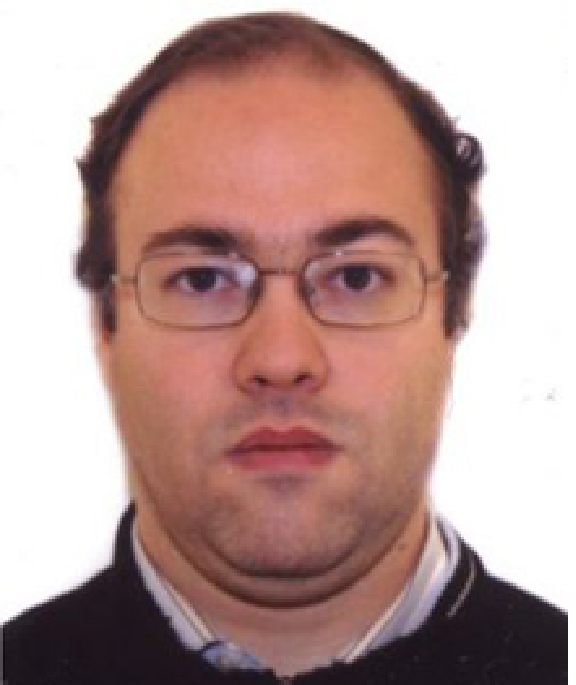}}]%
	{Giacomo Verticale} is assistant professor at Politecnico di Milano, Italy. He obtained his PhD in Telecommunication Engineering in year 2003 from Politecnico di Milano defending a thesis on the performance of packet transmission in UMTS. In the years 1999-2001 he was with the research center CEFRIEL, working on the Voice-over-IP and ADSL technologies. He was involved in several European research projects advancing the Internet technology. His current interests focus on the security issues of the Smart Grid and on NFV.
\end{IEEEbiography} 
\end{document}